\documentclass[12pt]{article}
\usepackage{amsmath, mathrsfs, amssymb}
\usepackage{graphicx}
\usepackage{subcaption}
\usepackage[round]{natbib}
\newtheorem{theorem}{Theorem}

\newtheorem{corollary}{Corollary}
\newtheorem{lemma}{Lemma}

\newtheorem{remark}{Remark}

\newtheorem{definition}{Definition}
\newtheorem{assumption}{Assumption}

\usepackage{bm}
\usepackage{booktabs}
\usepackage{algorithm}
\usepackage{algorithmic}
\renewcommand{\algorithmicrequire}{\textbf{Input:}}    \renewcommand{\algorithmicensure}{\textbf{Output:}}
\usepackage{color}
\usepackage{verbatim}%????
\usepackage{makecell}
\usepackage{enumerate}
\usepackage{threeparttable}
\usepackage{multirow}
\usepackage{hyperref}
\usepackage{setspace}
\usepackage{fontawesome}
\usepackage{accents}
\usepackage{enumitem}

 \hypersetup{
     colorlinks=true,
     linkcolor=cyan,
     filecolor=blue,
     citecolor = darkgreen,
     urlcolor=cyan,
     }
\newcommand{\tA}{{\tilde A}}

\newcommand{\MP}{{\mathbb P}}
\newcommand{\ME}{{\mathbb E}}

\newcommand{\mybar}[1]{\accentset{\rule{0.65em}{0.5pt}}{#1}}

\newcommand*{\QEDA}{\hfill\ensuremath{\blacksquare}}

\newcommand{\blind}{0}

\definecolor{darkgreen}{rgb}{0,0.6,0.2}
\allowdisplaybreaks[1]

\begin{document}

\def\spacingset#1{\renewcommand{\baselinestretch}%
{#1}\small\normalsize} \spacingset{1}

%%%%%%%%%%%%%%%%%%%%%%%%%%%%%%%%%%%%%%%%%%%%%%%%%%%%%%%%%%%%%%%%%%%%%%%%%%%%%%

\if0\blind
{
  \title{\bf Randomized Spectral Clustering for Large-Scale Multi-Layer Networks }
  \author{Wenqing Su$^1$, Xiao Guo$^2$, Xiangyu Chang$^3$, Ying Yang$^1$ \hspace{.2cm}  \\
  $^1$ Department of Mathematical Sciences, Tsinghua University, China  \\
    $^2$ School of Mathematics, Northwest University, China\\
    $^3$ School of Management, Xi'an Jiaotong University, China}
  \date{}
  \maketitle
} \fi

\spacingset{1.5} % DON'T change the spacing!

\bigskip
\begin{abstract}
Large-scale multi-layer networks with large numbers of nodes, edges, and layers arise  across various domains, which poses a great computational challenge for the downstream analysis. In this paper, we develop an efficient randomized spectral clustering algorithm for community detection of multi-layer networks. We first utilize the random sampling strategy to sparsify the adjacency matrix of each layer. Then we use the random projection strategy to accelerate the eigen-decomposition of the sum-of-squared sparsified adjacency matrices of all layers. The communities are finally obtained via the $k$-means of the eigenvectors. The algorithm not only has low time complexity but also saves the storage space. Theoretically, we study the misclassification error rate of the proposed algorithm under the multi-layer stochastic block models, which shows that the randomization does not deteriorate the error bound under certain conditions. Numerical studies on multi-layer networks with millions of nodes show the superior efficiency of the proposed algorithm, which achieves clustering results rapidly. A new R package called \texttt{MLRclust} is developed and made available to the public.

\end{abstract}
\vfill

\section{Introduction}\label{sec::intro}
Multi-layer networks have gained considerable attention in modern data analysis due to their ability to incorporate multiple relationships among the same set of entities, where the nodes are the entities of interest and each layer represents a different type of relationship. The multi-layer networks provide more information and insights than the single-layer network does \citep{kivela2014multilayer, boccaletti2014structure,bianconi2018multilayer}, and they have been widely studied and analyzed in various scientific fields including gene co-expression \citep{liu2018global}, social science \citep{omodei2015characterizing}, and neuroimaging \citep{paul2020random}, among others. Due to the advancements in computing and measurement technologies, the size of the multi-layer networks---nodes, edges and layers---are all increasing, making the analysis of multi-layer networks more challenging.

Community detection is a fundamental task in multi-layer network analysis and has seen rapid developments in recent years \citep{han2015consistent, matias2017statistical, pensky2019dynamic, paul2020spectral, lei2020consistent, wang2021fast, arroyo2021inference, jing2021community, lei2020bias}, especially within the framework of multi-layer Stochastic Block Models (SBMs) \citep{han2015consistent}. Unless otherwise specified, the communities are assumed to be consensus across layers. Specifically, the spectral clustering-based methods become popular due to the computational tractability. \cite{paul2020spectral} considered several spectral clustering-based methods including the spectral clustering of average adjacency matrices corresponding to all layers and the co-regularized spectral clustering. \cite{arroyo2021inference} developed a joint spectral embedding method of adjacency matrices, where the eigenspace of each network layer is properly aggregated to obtain the common eigenspace. \cite{jing2021community} developed a tensor-based spectral clustering method to detect the communities where there exist several classes of communities among layers. \cite{lei2020bias} proposed a spectral clustering method based on the bias-adjusted sum of squared adjacency matrices. Other spectral clustering-based methods include \cite{bhattacharyya2018spectral, pensky2019spectral, noroozi2022sparse, su2024spectral}, among others. Despite the tractability of the aforementioned methods, they are computationally expensive when the network size increases. Take the simple spectral clustering method of the average adjacency matrices as an example. The partial eigen-decomposition is burdensome when the number of nodes or the number of non-zero entries in the average adjacency matrices is large, where the latter would happen even if every single network is sparse. In this regard, the computational issues in multi-layer networks are more severe than those in single networks.

Randomization has been a useful probabilistic technique for solving large-scale linear algebra problems with high precision, beyond the computational capabilities of classical methods \citep{mahoney2009cur, woodruff2014sketching, martinsson2020randomized}. Generally, it utilizes a certain degree of randomness,  such as random projection and random sampling, to construct a \emph{sketch} of the full dataset, which retains the main information of the original dataset but has a much lower dimensionality or increased sparsity. Hence, by using the sketched dataset, the computational cost is largely  reduced. Randomization techniques have been thoroughly studied in linear regression and low rank matrix approximation problems; see \cite{drineas2012fast, raskutti2016statistical, ahfock2021statistical,halko2011finding}, among others. In the context of community detection, several works study randomized algorithms in the \emph{single} network set-up; see \cite{zhang2022randomized,guo2020randomized, deng2024subsampling}, for example. However, as mentioned earlier, the computational issues in multi-layer networks are much more severe. For the community detection of multi-layer networks, an efficient algorithm using randomization techniques is imminent and essential.

In this paper, we study the efficient community detection of large-scale multi-layer networks via randomization. We use the method in \citet{lei2020bias}, i.e., spectral clustering of the debiased sum of squared adjacency matrices, as the basis of our algorithm. The reason for choosing this method lies in the following aspects. First, in the real multi-layer networks, some layers are assortative mixing (nodes in the same community are more likely to link than nodes from distinct communities) and others are disassortative mixing (the opposite of assortative mixing), the squaring operator in the method of \citet{lei2020bias} is well accommodated to this unknown \emph{layer identity} problem. By contrast, aggregating the adjacency matrix without squaring can result in community cancellation. Second, the method is theoretically sound. It nearly attains optimal network density threshold when restricted to polynomial-time algorithms in terms of community recovery and detection \citep{lei2023computational}.

The main contributions of this paper are three-folds. First, we propose an efficient randomized spectral algorithm for detecting the communities of multi-layer networks. In particular, we first use the random sampling strategy to sparsify the adjacency matrices of all layers. Then, we use the random projection strategy to accelerate the eigen-decomposition of the sum of squared \emph{sparsified} adjacency matrices. Notably, we utilize the randomized block Krylov subspace \citep{musco2015randomized} in the approximation of eigenspace, which can attain better accuracy under the same number of power iterations than the subspace method  \citep{halko2011finding} used in \citet{guo2020randomized,zhang2022randomized}. The proposed randomized algorithm greatly reduces both the time complexity and space complexity compared to non-randomized methods. Second, we study the clustering performance of the resulting algorithms under the multi-layer SBM. It turns out that under certain conditions, the misclassification rate is the same as that of the corresponding algorithm without randomization. The algorithm and theory are both extended to the \emph{co-clustering} \citep{rohe2016co} of multi-layer \emph{directed} networks. Last but not least, to facilitate the implementation of the algorithm, we develop an R package called \texttt{MLRClust}\footnote{\url{https://github.com/WenqingSu/MLRClust}}. It turns out that the proposed algorithm is capable of processing large-scale multi-layer networks with millions of nodes in less than one minute. By contrast, the non-randomized counterpart algorithm, where the partial eigen-decomposition of the sum of squared original adjacency matrix is implemented by the fast iterative method \citep{baglama2005augmented}, fails when the number of nodes exceeds approximately twenty thousand in our experiment.

The remainder of the paper is organized as follows. Section \ref{sec::prob} provides the preliminaries. Sections \ref{sec::rand} and \ref{sec::thm} develop the randomized spectral clustering algorithm and provide its theoretical property, respectively. Sections \ref{sec::sims} and \ref{sec::real} include the simulation and real data experiments, respectively. Section \ref{sec::con} concludes the paper. In the Appendix, we provide the technical proofs. The extension of the method and theory to multi-layer stochastic co-block models for co-clustering is also included.

\section{Preliminaries}\label{sec::prob}
In this section, we introduce the widely used multi-layer SBMs and the spectral clustering algorithm for the purpose of developing the randomized algorithm. The following notation is used throughout the paper.

\emph{Notation:} We use $[n]$ to denote the set $\{1,...,n\}$. 
For a matrix $A\in \mathbb R^{n\times n}$, $\|A\|_0$ and $\|A\|_{\rm F}$ denote the number of its non-zero elements and its Frobenius norm, respectively. $\|\cdot\|_2$ denotes the spectral norm of a matrix or the Euclidean norm of a vector. $\|A\|_{2, \infty}$ denotes the maximum row-wise $l_2$ norm. For two sequences $f_n$ and $g_n$, $f_n = o(g_n)$ denotes that $f_n / g_n \rightarrow 0$ as $n$ goes to infinity. We write $f_n \lesssim g_n$ or $f_n = O(g_n)$ if there exists some constant $c>0$ such that $f_n \leq cg_n$, and write $f_n \asymp g_n$ if both $g_n \lesssim f_n$ and $f_n \lesssim g_n$ hold. 

We consider the following multi-layer SBM \citep{han2015consistent,paul2016consistent,barbillon2017stochastic} for generating the multi-layer network data. Suppose the multi-layer network has $L$ layers and $n$ nodes. For $l\in [L]$, $A_l\in \{0,1\}^{n\times n}$ corresponds to the network adjacency matrix of layer $l$, which is a symmetric matrix. We assume that each layer of network shares common community structure but has possibly different link probabilities. Specifically, for each $l$, $A_l$ is generated independently as follows,
\begin{equation}\label{msbm}
A_{l,ij}\sim_{i.i.d.}{\rm Bernoulli}(B_{l,g_ig_j})\; {\rm if} \; i<j \quad {\rm and} \quad A_{l,ij}=0\; {\rm if}\; i=j,
\end{equation}
where $g_i\in\{1,...,K\}$ denotes the community assignments of node $i$ for $i\in [n]$, and $B_l\in [0,1]^{K\times K}$ denotes the block probability matrix of layer $l$. Here, $K$ is the number of communities, for which we assume to be fixed throughout the paper. The population of $A_l$ can be formulated as $P_l:=\Theta B_l\Theta$, where $\Theta\in \{0,1\}^{n\times K}$ is the membership matrix whose $i$th row corresponds to node $i$ and has exactly one 1 on the $g_i$'s column and $K-1$ 0's elsewhere.

To estimate the communities in the multi-layer SBM, the spectral clustering method performed on the following aggregated matrix $M$ has become popular and studied by \citet{lei2020bias,lei2023computational,su2024spectral,xie2024bias} among others, 
\begin{equation}
\label{eq:sosa}
  M:=\frac{1}{Ln}\sum_{l=1}^L (A_l^2-D_l),
\end{equation}
where $D_l$ is a diagonal matrix consisting of the degree of $A_l$ with $D_{l,ii}=\sum_{j=1}^n A_{l,ij}$. Instead of summing the matrices $A_l$ directly, squaring each $A_l$ before the summation could adapt to the heterogeneity that some layers of networks may be densely connected \emph{within} communities while other layers may be densely connected \emph{across} communities. Subtracting $D_l$ is to reduce the statistical bias caused by squaring. The spectral clustering on $M$ consists of two steps, the eigen-decomposition is first performed on $M$ to obtain the top $K$ eigenvectors of $M$, and then the $k$-means clustering is conducted on them to estimate the communities. It turns out that $M$ can well-approximate its population counterpart
\begin{equation}\label{Pi}
{\Pi}:=\frac{1}{L}\sum_{l=1}^L \frac{P_l^2}{n}:=\frac{1}{L}\sum_{l=1}^L {\Pi}_l.
\end{equation}

Though the spectral clustering algorithm is conceptually easy, it is actually computationally inefficient from the following aspects. First, squaring all $A_l$ generally requires $O(Ln \max_l\|A_l\|_0)$ time, which is costly when some of $L,n$ and $\|A\|_0$ are large. Second, after the summation of $A_l^2$ in \eqref{eq:sosa}, $M$ would be a dense matrix. The full eigen-decomposition of $M$ would thus generally takes $O(n^3)$ time. The iterative algorithm for partial eigen-decomposition takes $O(n^2T)$ time, which is also time costly when $n$ is large, where $T$ denotes the number of iterations.  We will see in the next section how the randomized sketching techniques could be used to speed up the spectral clustering algorithm.

\section{Randomized spectral clustering for multi-layer SBMs}\label{sec::rand}
In this section, we introduce random sampling and random projection strategies, and provide the randomized spectral clustering algorithm designed for multi-layer SBMs.
\subsection{Randomization}
\paragraph{Random sampling.} To reduce the computational burden of squaring $A_l$, we employ the following \emph{random sampling} strategy to construct sparsified matrices $\tA_l$. Specifically, for each pair $1 \leq i < j \leq n$,
\begin{equation}\label{rs}
	{\tA}_{l,ij}=\begin{cases}
	\frac{A_{l,ij}}{p}, & \mbox{if }\; (i,j) {\mbox{ is selected},} \\
	0,& \mbox{if } \;(i,j) {\mbox{ is not selected}},
	\end{cases}
\end{equation}
and $\tA_{l,ij} = \tA_{l,ji}$ for each $i > j$. Here, $A_{l,ij}$ is divided by $p$ in order to satisfy that $\mathbb E(\tA_l)=\mathbb E(A_l)$. Hopefully, when the sampling probability $p$ is not too small, the characteristics of $\tA_l$ would be close to those of $A_l$ without sacrificing too much accuracy.

\paragraph{Time complexity.} The time complexity of $\sum_{l=1}^L\tA_l^2$ would generally be reduced to $O(Ln \max_l\|\tA_l\|_0)$, which can be even smaller in practice, depending on the structure of $\tA_l$'s. The time complexity for random sampling is only $O(L \max_l\|A_l\|_0)$.

\paragraph{Random projection.} One can imagine that $\sum_{l=1}^L A_l^2$ is biased from its population counterpart $\sum_{l=1}^L P_l^2$. But before carefully calculating the bias, for the moment, we assume that we have obtained a bias-corrected counterpart of $\frac{1}{Ln}\sum_{l=1}^L \tA_l^2$, denoted by $\mybar{M}$. We now focus on illustrating how we can use the \emph{random projection} strategy to accelerate the eigen-decomposition of $\mybar{M}$. The general idea is to use the following randomized block Krylov methods \citep{musco2015randomized, li2021randomized} developed by the computer science community. First, an orthogonal basis $Q\in \mathbb R^{n\times (q+1)K} $ is obtained via the QR decomposition of a compressed matrix $$G:=[\mybar{M}^{1}\Omega, \mybar{M}^{3}\Omega, \ldots, \mybar{M}^{2q+1}\Omega],$$ where $\Omega$ is an $n\times K$ random test matrix whose entries could be i.i.d. standard Gaussian and Rademacher, among others. Then the eigenvectors of $\mybar{M}$ could be approximated by $QU_K$, where $U_K$ consists of the top $K$ eigenvector of the small matrix $Q^T \mybar{M} Q$. In essence, the random projection here aims to identify an orthogonal basis $Q$ that facilitates the construction of a low-rank approximation of  $\mybar{M}$ such that $$\mybar{M} \approx QQ^T \mybar{M} QQ^T.$$

\paragraph{Time complexity.} The time complexity of the whole procedure is dominated by the matrix multiplications for forming $G$, which is $O((2q+1)n^2K)$. And thus the computational burden of the eigen-decomposition of $\mybar{M}$ is largely reduced.

\paragraph{Discussions.} The practical time complexity of the {random projection} procedure is much lower, as we do not need to compute the dense matrix $\mybar{M}$ when forming $G$. Instead, the matrix multiplications for forming $G$ are accomplished by left-multiplying the random test matrix $\Omega$ by the sparsified adjacency matrices, which have a time complexity of $O(LqK\max_l\|\tA_l\|_0)$. Notably, $\tA_l$ is relatively sparse, and the number of layers $L$ is typically much less than $n$ in large-scale multi-layer networks, thus this complexity is less than in cases where $\mybar{M}$ is computed directly, which takes $O((2q+1) n^2K)$ time.

More importantly, this methodology not only avoids the extensive computational requirements associated with computing $\mybar{M}$ but also mitigates the challenges of space complexity. This is crucial because the space complexity of a dense matrix is $O(n^2)$, potentially leading to memory overflows in scenarios where $n$ is relatively large, say networks with millions of nodes.

It is worth mentioning that the random projection can be made more efficient with parallel and distributed computing and it has low communication costs due to requiring only a few passes over the input matrix.

% As a result, the randomization procedure will require a total time of $O(LqK\max_l\|\tA_l\|_0)$ + $O(Ln\max_l\|\tA_l\|_0)$  and will avoid the need to store dense large square matrices, which significantly reduces the computational burden of spectral clustering.
% \xiaoc{The total time would be even reduced if considering distributed computing}
% \xiaoc{Can we add a table to compare the time complexity with the non-randomized algorithm.}

%The general idea is to find an orthogonal basis $Q\in \mathbb R^{n\times K} $ such that $$\mybar{M} \approx QQ^T \mybar{M} QQ^T.$$ Then the eigenvector of $\mybar{M}$ could be approximated by $QU_s$, where $U_s$ consists of the top $K$ eigenvector of the small matrix $Q^T \mybar{M} Q$. The remaining procedure is to build the orthogonal basis $Q$, which can be regarded as the low-rank approximation of the column space of $\mybar{M}$. Actually, $Q$ is obtained via the QR decomposition of a compressed matrix $G:=[\mybar{M}^{1}\Omega, \mybar{M}^{3}\Omega, \ldots, \mybar{M}^{2q+1}\Omega]$, where $\Omega$ is an $n\times K$ random test matrix whose entries could be i.i.d. standard Gaussian and Rademacher, among others. The time complexity of the whole procedure is dominated by the matrix multiplications for forming $G$, which is $O((2q+1)n^2K)$. And thus the computational burden of the eigen-decomposition of $\mybar{M}$ is largely reduced.

\begin{remark}
There have been several prior works on spectral clustering using randomized methods, see \citet{yan2009fast,sakai2009fast,sinha2018k,tremblay2016compressive,wang2019scalable,tremblay2020approximating}, among others. These works focus on general dataset and convergence theory. 
Different from these works, we develop a randomization method for clustering multi-layer network data. The statistical theory explains how randomization influences the misclassification error rate. On the other hand, the randomization has also been studied under single statistical network models \citep{li2020network,zhang2022randomized,deng2024subsampling}. In this work, we turn to focus on the multi-layer networks, which are more computationally challenging than single networks. We tailor the randomization techniques for the spectral clustering on the sum of squared matrix and study the misclassification rate of the proposed algorithm under the multi-layer SBM and its directed variant. 
\end{remark}

\subsection{Bias-adjustment}\label{bias adjustment}
As mentioned earlier, $\sum_{l=1}^L\tA_l^2$ would introduce bias compared to its population counterpart $\sum_{l=1}^L P_l^2$. We now proceed to debias.

It is easy to see that, ignoring the diagonal elements, $\tA_l$ is an unbiased estimator for $P_l$. Our main effort is to show how the diagonal elements of $\tA_l^2$ would cause bias for $P_l^2$.
Denote the non-diagonal part of $P_l$ by $\bar{P}_l:=P_l-{\rm diag}(P_l)$ and let $X_l:=\tA_l-\bar{P}_l$. Then,
\begin{align*}
\tA_l^2&=(X_l+\bar{P}_l)^2=X_l^2+\bar{P}_l^2+X_l \bar{P}_l+\bar{P}_l X_l.
\end{align*}
By plugging in
$\bar{P}_l^2=(P_l-{\rm diag}(P_l))^2$,
we further have
\begin{align}
\label{composion2}
\tA_l^2-P_l^2&=E_{l,0}+E_{l,1}+E_{l,2}+E_{l,3}
\end{align}
with
\begin{align*}
E_{l,0}&:={-P_l{\rm diag}(P_l)-{\rm diag}(P_l)\cdot P_l +({\rm diag}(P_l))^2},\\
E_{l,1}&:=X_l(P_l-{\rm diag}(P_l))+ (P_l-{\rm diag}(P_l))X_l,\\
E_{l,2}&:=X_l^2-{\rm diag}(X_l^2),\\
E_{l,3}&:={\rm diag}(X_l^2).
\end{align*}
The first three terms $E_{l,0}$, $E_{l,1}$, and $E_{l,2}$ turn out to be small. We thus focus on the fourth term $E_{l,3}$. Note that
\begin{align}
\label{e3com}
(E_{l,3})_{ii}&=\sum_{j=1}^n X_{l,ij}^2=\sum_{j=1,j\neq i}^n (\tA_{l,ij}-P_{l,ij})^2\nonumber\\
&=\sum_{j=1}^n \mathbb I(\tA_{l,ij}=0)P_{l,ij}^2+\sum_{j=1}^n \mathbb I(\tA_{l,ij}=1/p)({1}/{p}-P_{l,ij})^2\nonumber\\
&\leq n\cdot \max_{l,ij}P^2_{l,ij}+\tilde{d}_{l,i}\cdot \frac{1}{p^2},
\end{align}
where $\tilde{d}_{l,i}:=\sum_{j=1}^n\mathbb I(\tA_{l,ij}=1/p)$ denotes the number of edges of node $i$ in the $l$-th network. Simple calculation shows that the expectation of the second term in \eqref{e3com} is $\mathbb E(\tilde{d}_{l,i})/p^2=\sum_{j=1}^nP_{l,ij}/ p \leq n \max_{l,ij}P_{l,ij}/p$. Hence, $\tilde{d}_{l,i}/{p^2}$ dominates $n\cdot \max_{l,ij}P^2_{l,ij}$ since $1/p>1$.

In light of the above calculation, we derive the following debiased estimator $\mybar{M}$,
\begin{equation*}\label{mbar}
	\mybar{M}:=\frac{1}{L}\sum_{l=1}^L \mybar{M}_l
\end{equation*}
with
\begin{equation}\label{dm}
	\mybar{M}_l:= \frac{\tA_l^2}{n}-\frac{\tilde{D}_{l}}{np^2},
\end{equation}
where $\tilde{D}_l$ is a diagonal matrix with $\tilde{D}_{l,ii}=\tilde{d}_{l,i}$. $\mybar{M}$ can then be used as the input of the random projection strategy for fast eigen-decomposition.

\begin{remark}
Note that the bias-adjustment for the sum of squared \emph{original} adjacency matrices is first developed in \citet{lei2020bias} and further refined by \citet{xie2024bias}. Equation \eqref{dm} is in a similar spirit of \citet{lei2020bias} except that we deal with the matrix after random sampling.  
\end{remark}

\subsection{Algorithm}
Based on the above discussions, we propose the \textbf{r}andomized \textbf{s}pectral \textbf{c}lustering algorithm (\texttt{RSC}) for multi-layer SBMs, summarized in Algorithm \ref{rsc}.

For clarity, we retain the step of forming the matrix $\mybar{M}$ in the algorithm; however, as mentioned earlier, this step can be omitted in practical implementations. Typically, matrix $G$ can be efficiently constructed directly by left-multiplying the random test matrix $\Omega$ by the sparsified adjacency matrices, offering significant computational savings in both time and space complexity.

\begin{algorithm}[!htbp]
\small
\renewcommand{\algorithmicrequire}{\textbf{Input:}}
\renewcommand\algorithmicensure {\textbf{Output:} }
\caption{\textbf{R}andomized \textbf{s}pectral \textbf{c}lustering for multi-layer SBMs (\texttt{RSC})}
\label{rsc}
\begin{algorithmic}[1]
\STATE \textbf{Input:} Multi-layer network adjacency matrices $\{A_l\}_{l=1}^L$, number of clusters $K$, sampling probability $p$, and power parameter $q$.  \\
\FOR{$l = 1$ to $L$}
\STATE \textbf{Random sampling:} Randomly sample $A_l$'s elements to obtain a sparsified matrix $\tA_l$ according to \eqref{rs};\\
\STATE \textbf{Bias-correction:} Squaring $\tA_l$'s and construct the debiased estimator $\mybar{M}_l$ according to \eqref{dm}.
\ENDFOR
\STATE \textbf{Random-projection-based eigen-decomposition:} \\
\STATE \quad Form $\mybar{M}$ by $\mybar{M}:=\sum_{l=1}^L \mybar{M}_l/L$;
\STATE \quad Draw an $n \times K$ random test matrix $\Omega$;\\
\STATE \quad Form the matrix $G=[\mybar{M}^{1}\Omega, \mybar{M}^{3}\Omega, \ldots, \mybar{M}^{2q+1}\Omega]$;\\
\STATE \quad Construct $Q \in \mathbb R^{n\times (q + 1)K}$ via QR decomposition of $G$, $G=:QR$;\\
\STATE \quad Form $C=Q^T\mybar{M} Q \in \mathbb R^{(q + 1)K \times (q + 1)K}$;\\
\STATE \quad Set $\hat{U}:=Q{U}_K$ with $U_K$ being the top $K$ eigenvectors of the small matrix $C$.\\
\STATE \textbf{Clustering:} Conduct $k$-means clustering on $\hat{U}$ with cluster number being $K$. The resulting
clusters are formalized as the membership matrix $\hat{\Theta}$. \\
\STATE \textbf{Output:} The estimated membership matrix $\hat{\Theta}$.
\end{algorithmic}
\end{algorithm}

%\begin{algorithm}[!htbp]
%\small
%\renewcommand{\algorithmicrequire}{\textbf{Input:}}
%\renewcommand\algorithmicensure {\textbf{Output:} }
%\caption{\textbf{R}andomized \textbf{s}pectral \textbf{c}lustering for multi-layer networks (\texttt{RSC})}
%\label{rsc}
%\begin{algorithmic}[1]
%\STATE \textbf{Input:} Multi-layer network adjacency matrices $\{A_l\}_{l=1}^L$, number of clusters $K$, sampling probability $p$, and power parameter $q$.  \\
%\FOR{$l = 1$ to $L$}
%\STATE \textbf{Random sampling:} Randomly sample $A_l$'s elements to obtain a sparsified matrix $\tA_l$ ;\\
%\STATE \textbf{Bias-correction:} Squaring $\tA_l$'s and construct the debiased estimator $M_l$.
%\ENDFOR
%\STATE \textbf{Random-projection-based eigen-decomposition:} \\
%\STATE \quad Draw an $n \times K$ random test matrix $\Omega$;\\
%\STATE \quad Form the matrix $G=[M\Omega, (MM^{\T})M\Omega, \ldots, (MM^{\T})^qM\Omega]$;\\
%\STATE \quad Construct $Q \in \mathbb R^{n\times (q+1)K}$ via QR decomposition of $G$, $G=:QR$;\\
%\STATE \quad Form $C=Q^T\mybar{M} Q \in \mathbb R^{(q+1)K \times (q+1)K}$;\\
%\STATE \quad Set $U:=Q{\tilde U}_K$ with $\tilde U_K$ being the $K$ largest eigenvectors of the small matrix $C$.\\
%\STATE \textbf{Clustering:} Conduct $k$-means clustering on $U$ with cluster number being $K$. 
%\end{algorithmic}
%\end{algorithm}
%

\section{Consistency}\label{sec::thm}
In this section, we study how the randomization affects the clustering performance of \texttt{RSC}. We measure the clustering performance using the misclassification rate defined as
\begin{equation*}\label{misclustering rate}
	\mathcal{L}(\Theta,\widehat{\Theta}) = \min_{\Psi \in \bm{\Psi}_{K}} \frac{1}{n}\|\widehat{\Theta}\Psi - \Theta\|_0,
\end{equation*} 
where $\widehat{\Theta}\in \{0,1\}^{n\times K}$ and $\Theta$ correspond to the estimated and true membership matrices, and $\bm{\Psi}_{K}$ is the set of all $K \times K$ permutation matrices.

The following assumptions are needed to bound the misclassification rate of \texttt{RSC}.
%\begin{assumption}[Sparsity]
%\label{sparse}
%The block probability matrix $B_l =\rho B_{l,0}$, where $\rho=o(1)$ and $B_{l,0}$ is a $K\times K$ symmetric matrix with constant entries in $[0,1]$.
%\end{assumption}

\begin{assumption}[Balanced communities]
\label{balcom}
For $k\in[K]$, $n_k$ denotes the number of nodes within the $k$-th community. The community is balanced such that for any $k\in[K]$ and some constants $c$ and $C$,
\begin{equation*}
c\frac{n}{K}\leq n_k\leq C\frac{n}{K}.
\end{equation*}
\end{assumption}

\begin{assumption}[Eigen-gap and full rank]
\label{rank}
$\sum_{l=1}^L B_{l,0}^2$ has full rank $K$ with $$\sigma_{\rm min}(\sum_{l=1}^L B_{l,0}^2)\geq cL,$$ where $\sigma_{\rm min}(\cdot)$ denotes the minimum eigenvalue of a matrix.
\end{assumption}

{Assumption \ref{balcom} requires that the communities are balanced, which is a common assumption for notational simplicity in the derived misclassification rate. Assumption \ref{rank} requires the linear growth of the aggregated squared block probability matrices in terms of its minimum eigenvalue; similar conditions are also imposed in \cite{lei2020bias}. Note that when $B_l$'s are identical and positive-definite across layers with two parameters (i.e., all the diagonal and non-diagonal entries are the same, respectively), Assumption \ref{rank} holds naturally. For analytical simplicity, we assume that $\sum_{l=1}^L B_{l,0}^2$ is of full rank, this condition can be relaxed to a rank-deficient scenario, where only a linear growth of the minimum \emph{non-zero} eigenvalue of the aggregated squared block probability matrices is required \citep{su2024spectral}.}

The next theorem provides the bound for the misclassification rate of \texttt{RSC} under two sparsity regimes. 

\begin{theorem}\label{thm:mr}
	Suppose Assumptions \ref{balcom} and \ref{rank} hold and the power parameter $q \asymp \log n$, the following results hold respectively for two sparsity regimes.
\begin{enumerate}[label=(\roman*)] 
\item  If
\begin{equation}\label{sparse-condition}
	\frac{\log (L+n)}{pnL^{1/2}}\lesssim\rho\lesssim \frac{1}{pn},
\end{equation}
then with probability larger than $1-O((L+n)^{- c_1}) - 2e^{-c_2K}$ for some positive constants $c_1$ and $c_2$,
\begin{align}\label{mr-sparse-bound}
	\mathcal{L}(\Theta,\widehat{\Theta})\lesssim\max \{\frac{\log^{2}(L+n)}{p^{3}Ln^{2}\rho^2},\; \frac{1}{n^2}\}.
\end{align}
\item If
\begin{equation}\label{dense-condition}
	\rho\gtrsim \frac{\log (L+n)}{pn},
\end{equation}
then with probability larger than $1-O((L+n)^{- c_3}) - 2e^{-c_4K}$ for some positive constants $c_3$ and $c_4$,
\begin{align}\label{mr-dense-bound}
	\mathcal{L}(\Theta,\widehat{\Theta})\lesssim\max \{\frac{\log(L+n)}{pLn\rho},\;\frac{1}{n^2}\}. \end{align}
\end{enumerate}
\end{theorem}

\begin{remark}
In the regime (\ref{sparse-condition}), the lower bound of $\rho$ depends on $L^{-1/2}$, which means that the sparsity level $\rho$ in each single layer can be small provided that the number of layers $L$ is large, indicating the benefit of combining $L$ layers. This situation (i.e., lower bound of $\rho$ inversely proportional to $L$) is generally more challenging in the statistical estimation of multi-layer SBMs (e.g., \citet{lei2020bias,jing2021community}). In this work, we address not only the statistical estimation but also the computational challenges associated with large-scale SBMs. Therefore, we also examine the regime (\ref{dense-condition}) with a higher sparsity level, which is more challenging than (\ref{sparse-condition}) from the computational aspect. 
\end{remark}

\begin{remark}
Under the regime that $\log^{1/2}(L+n)/(L^{1/2}n)\lesssim\rho\lesssim 1/n$, which corresponds to our regime \eqref{sparse-condition} for fixed $p$ up to log terms, \citet{lei2020bias} derived the bound $\max\{\log(L+n)/(Ln^2\rho^2),\;n^{-2}\}$ for misclassification rate. Hence, {\rm \eqref{mr-sparse-bound}} is of the same order with their bound for fixed $p$ up to log terms, though {\rm \eqref{mr-sparse-bound}} is generally not better than theirs due to the information loss of random sampling. 

{Under the regime $n\rho \gtrsim \log n$, which corresponds to our regime \eqref{dense-condition} for fixed $p$ up to log terms, \citet{arroyo2021inference} derived the bound $\max\{1/(Ln\rho), n^{-2}\rho^{-2}\}$ for misclassification rate. Under $Ln\rho \gtrsim \log^4 n$, \citet{jing2021community} derived the bound $\log n /(Ln\rho)$. Hence, our bound under the regime \eqref{dense-condition} is of the same order as their bounds (under the sparsity regime \eqref{dense-condition}) for a fixed $p$ up to logarithmic terms. Note that \citet{jing2021community} utilized tensor-based methods for clustering, which yields tighter bounds compared to (\ref{mr-sparse-bound}) obtained with the summary statistics-based approach in this work.}
\end{remark}

\begin{remark}
We make use of the randomization results \citep{musco2015randomized} and decoupling techniques \citep{de1995decoupling} to derive the bound in Theorem \ref{thm:mr}. Note that due to randomization, directly applying the concentration tools developed in \cite{lei2020bias} does not yield a sharp bound in the sparse regime.
\end{remark}

% \begin{remark} 
% To simplify our theoretical bounds, we impose a full rank constraint on $\sum_{l=1}^L B_{l}^2$. Actually, this constraint can be relaxed to accommodate rank-deficient situations, and similar upper bounds on misclassification rates can be derived. For details, see \cite{su2024spectral}.
% \end{remark}

It is worth mentioning that the method and theory can be extended to the co-clustering of multi-layer directed networks under the  multi-layer Stochastic co-Block Models (ScBMs); see the Appendix for details. For simplicity, we will still refer to the randomized spectral co-clustering for multi-layer ScBMs as RSC.

\section{Simulations}\label{sec::sims}
In this section, we compare the finite sample performance of the proposed algorithm \texttt{RSC} with that of the original spectral clustering (\texttt{SC}) through simulations, to show that the accuracy of \texttt{RSC} is not much worse than \texttt{SC}.
We consider four model set-ups for the multi-layer networks: the first three models are based on the multi-layer SBM under three different scenarios; the fourth model corresponds to the multi-layer ScBM. 
\paragraph{\textbf{Model 1.}} Consider the $n$-node multi-layer SBM with $K = 3$ communities. The proportions of nodes in three communities are  $(0.3,0.4,0.3)$. We set $B_l = \rho B^{(1)}$ for  $l\in \{1, \ldots, L/2\}$ and $B_l = \rho B^{(2)}$ for $l\in \{L/2 + 1, \ldots, L\}$, with 
 \begin{equation*}
		B^{(1)} = U\begin{bmatrix} 1.5 & 0 & 0 \\ 0 & 0.2 & 0 \\ 0 & 0 & 0.4 \end{bmatrix}U^T \approx \begin{bmatrix}0.62 & 0.22 & 0.46 \\ 0.22 & 0.62 & 0.46 \\ 0.46 & 0.46 & 0.85 \end{bmatrix}
	\end{equation*} 
	and
	\begin{equation*}
		B^{(2)} = U\begin{bmatrix} 1.5 & 0 & 0 \\ 0 & 0.2 & 0 \\ 0 & 0 & -0.4 \end{bmatrix}U^T \approx \begin{bmatrix} 0.22 & 0.62 & 0.46 \\ 0.62 & 0.12 & 0.46 \\ 0.46 & 0.46 & 0.85 \end{bmatrix},
	\end{equation*} 
	where 
	\begin{equation}\label{U}
		U = \begin{bmatrix}
			1/2 & 1/2 & -\sqrt{2}/2 \\ 1/2 & 1/2 & \sqrt{2}/2 \\ \sqrt{2}/2 & -\sqrt{2}/2 & 0
		\end{bmatrix}. 	
	\end{equation}
	It is easy to see that $\sum_{l=1}^LB_l^2$ is of full rank. The networks are then generated from the multi-layer SBM given in \eqref{msbm}. This model was also studied in \cite{lei2020bias}.

\paragraph{\textbf{Model 2.}} In this model, we consider the case where $\sum_{l=1}^L B_l^2$ is rank-deficient. The basic set-up is identical to Model 1 except that we use the following matrix to define the block probability matrices.
 \begin{equation*}
		B^{(1)} = U\begin{bmatrix} 1.5 & 0 & 0 \\ 0 & 0.4 & 0 \\ 0 & 0 & 0 \end{bmatrix}U^T \approx \begin{bmatrix}0.66 & 0.22 & 0.44 \\ 0.22 & 0.46 & 0.58 \\ 0.44 & 0.58 & 0.78 \end{bmatrix}
	\end{equation*} 
	and
	\begin{equation*}
		B^{(2)} = U\begin{bmatrix} 1.5 & 0 & 0 \\ 0 & -0.4 & 0 \\ 0 & 0 & 0 \end{bmatrix}U^T \approx \begin{bmatrix} 0.09 & 0.53 & 0.62 \\ 0.53 & 0.29 & 0.48 \\ 0.62 & 0.48 & 0.72 \end{bmatrix},
	\end{equation*} 
	where 
	\begin{equation*}
		U = \begin{bmatrix}
			0.5 & 0.84 & -0.19 \\ 0.5 & -0.46 & -0.73 \\ 0.71 & -0.27 & 0.65
		\end{bmatrix}. 	
	\end{equation*}
	It is easy to see that $\sum_{l=1}^LB_l^2$ is rank-deficient and the rank is 2. We generate the adjacency matrices using the multi-layer SBM defined in \eqref{msbm}.

\paragraph{\textbf{Model 3.}} In previous models, we considered scenarios where direct summation of $B_l$ matrices would lead to the cancellation of some communities. We now consider a multi-layer SBM without this constraint. The model set-up is identical to Model 1, with the exception of 
	\begin{equation*}
		B^{(2)} = U\begin{bmatrix} 1 & 0 & 0 \\ 0 & 0.4 & 0 \\ 0 & 0 & 0.2 \end{bmatrix}U^T \approx \begin{bmatrix} 0.45 & 0.25 & 0.21 \\ 0.25 & 0.45 & 0.21 \\ 0.21 & 0.21 & 0.7 \end{bmatrix}.
	\end{equation*} 
The matrices $B^{(1)}$ and $U$ remain the same as those specified in Model 1. In this set-up, the direct sum does not lead to community cancellation.

\paragraph{\textbf{Model 4.}} In this model, we  consider the multi-layer ScBM (see Appendix \ref{appendix extension}), which is used for generating multi-layer directed networks. We set $B_l = \rho B^{(1)}$ for $l \in \{1,\ldots, L/2\}$, and $B_l = \rho B^{(2)}$ for $l \in \{L/2+1,\ldots, L\}$, with 
	\begin{equation*}
		B^{(1)} = U\begin{bmatrix} 1.5 & 0 & 0 \\ 0 & 0.2 & 0 \\ 0 & 0 & 0.4 \end{bmatrix}V^T \approx \begin{bmatrix}0.46 & 0.625 & 0.225 \\ 0.46 & 0.225 & 0.625 \\ 0.85 & 0.46 & 0.46 \end{bmatrix}
	\end{equation*} 
	and
	\begin{equation*}
		B^{(2)} = U\begin{bmatrix} 1.5 & 0 & 0 \\ 0 & 0.2 & 0 \\ 0 & 0 & -0.4 \end{bmatrix}V^T \approx \begin{bmatrix}0.46 & 0.225 & 0.625  \\ 0.46 & 0.625 & 0.225 \\ 0.85 & 0.46 & 0.46 \end{bmatrix},
	\end{equation*} 
	where $U$ is as defined in \eqref{U} and
	\begin{equation*}
		V = \begin{bmatrix}
			\sqrt{2}/2 & -\sqrt{2}/2 & 0 \\ 1/2 & 1/2 & -\sqrt{2}/2 \\ 1/2 & 1/2 & \sqrt{2}/2
		\end{bmatrix}.		
	\end{equation*}
We consider the $n$-node ScBM with $K_y = 3$ row communities and $K_z = 3$ column communities.
The proportions of nodes in three row (resp. column) communities are  $(0.3,0.4,0.3)$ (resp. $(0.4,0.3,0.3)$.

\paragraph{\textbf{Results.}}  For each model, we examine the misclassification rates as the number of nodes $n$, the number of layers $L$, and the sparsity $\rho$ vary, respectively, as shown below. 
\begin{enumerate}[label=(\roman*)] 
	\item Effect of $n$. $L = 20$, $\rho = 0.1$ and $n$ ranges from 200 to 2000.
	\item Effect of $L$. $n = 1000$, $\rho = 0.1$ and $L$ ranges from 2 to 50.
	\item Effect of $\rho$. $n = 1000$, $L = 20$ and $\rho$ ranges from 0.02 to 0.18.
\end{enumerate}
%In each parameter set-up we vary the sampling rate $p \in \{0.5, 0.7, 0.9\}$ and the power parameter $q \in \{2, 4, 6\}$. 
For each of the above three set-ups, we test different combinations of the sampling rate $p \in \{0.5, 0.7, 0.9\}$ and the power parameter $q \in \{2, 4, 6\}$.  % For computational convenience, we employ uniform sampling in the random sampling scheme. \xiaoc{What does uniform sampling means} 

\begin{figure*}[!htbp]
    \centering
    \begin{subfigure}{0.325\textwidth}
    	\includegraphics[width=\textwidth]{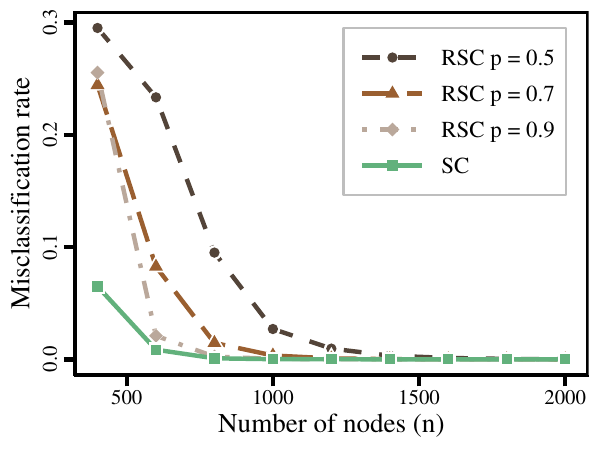}
    	\vspace{-0.13\textwidth}
        \caption{Effect of $n$, $q = 4$}
    \end{subfigure}
    \begin{subfigure}{0.325\textwidth}
        \includegraphics[width=\textwidth]{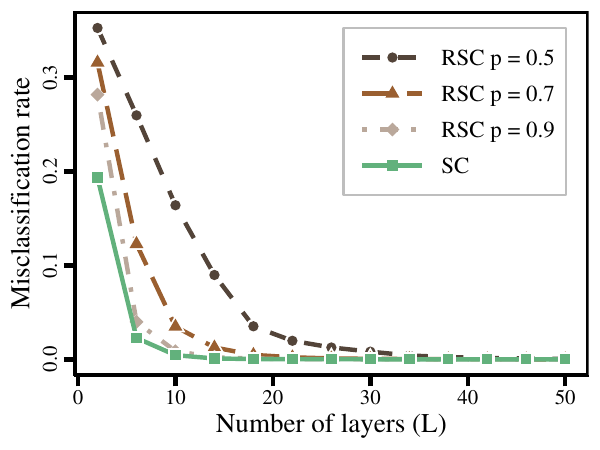}
    	\vspace{-0.13\textwidth}
        \caption{Effect of $L$, $q = 4$}
    \end{subfigure}
    \begin{subfigure}{0.325\textwidth}
        \includegraphics[width=\textwidth]{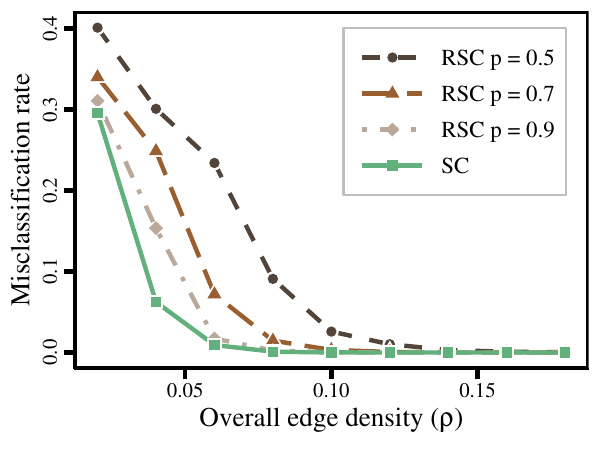}
    	\vspace{-0.13\textwidth}
        \caption{Effect of $\rho$, $q = 4$}
    \end{subfigure} \\
    \vspace{0.02\textwidth}
    \begin{subfigure}{0.325\textwidth}
    	\includegraphics[width=\textwidth]{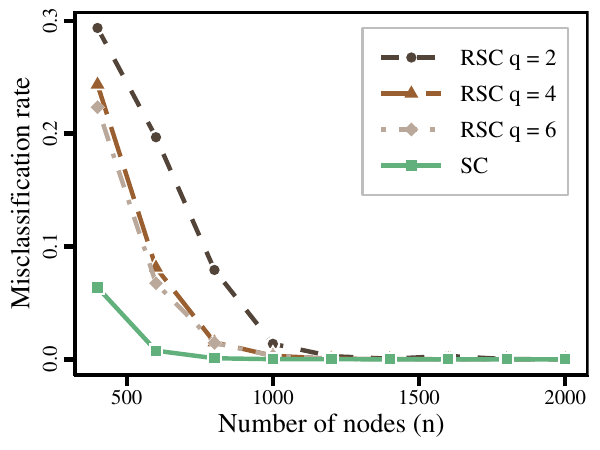}
    	\vspace{-0.13\textwidth}
        \caption{Effect of $n$, $p = 0.7$}
    \end{subfigure}
    \begin{subfigure}{0.325\textwidth}
        \includegraphics[width=\textwidth]{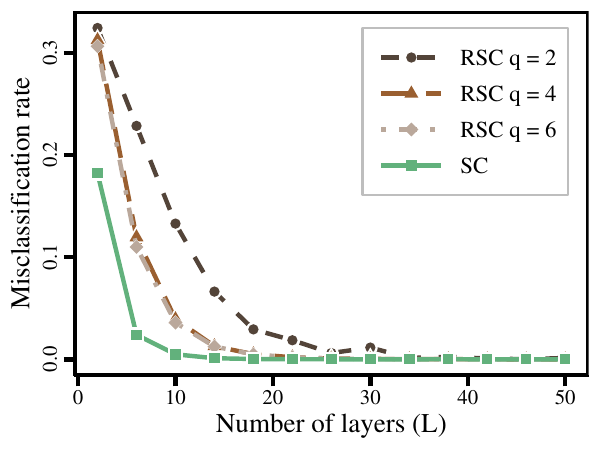}
    	\vspace{-0.13\textwidth}
        \caption{Effect of $L$, $p = 0.7$}
    \end{subfigure}
    \begin{subfigure}{0.325\textwidth}
        \includegraphics[width=\textwidth]{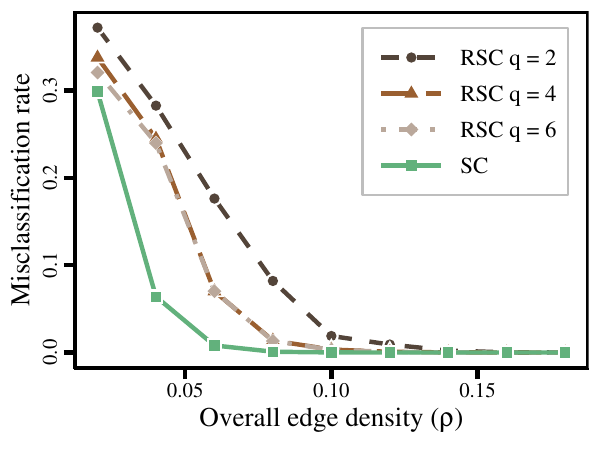}
    	\vspace{-0.13\textwidth}
        \caption{Effect of $\rho$, $p = 0.7$}
    \end{subfigure} 
    \caption{The misclassification rates of two methods under Model 1 with different parameters and hyperparameters.}
    \label{sim::model1}
\end{figure*}

\begin{figure*}[!htbp]
    \centering
    \begin{subfigure}{0.325\textwidth}
    	\includegraphics[width=\textwidth]{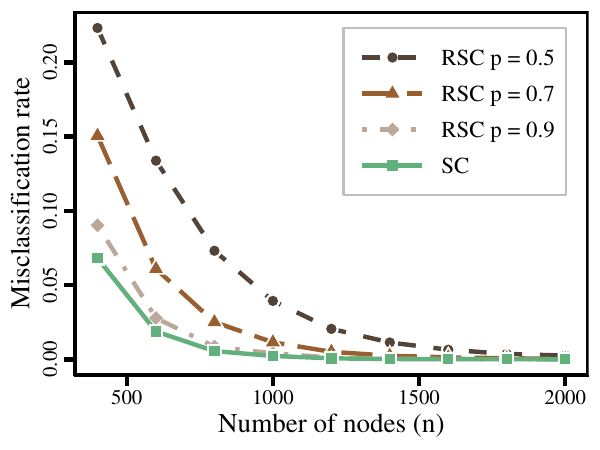}
        \caption{Effect of $n$, $q = 4$}
    \end{subfigure}
    \begin{subfigure}{0.325\textwidth}
        \includegraphics[width=\textwidth]{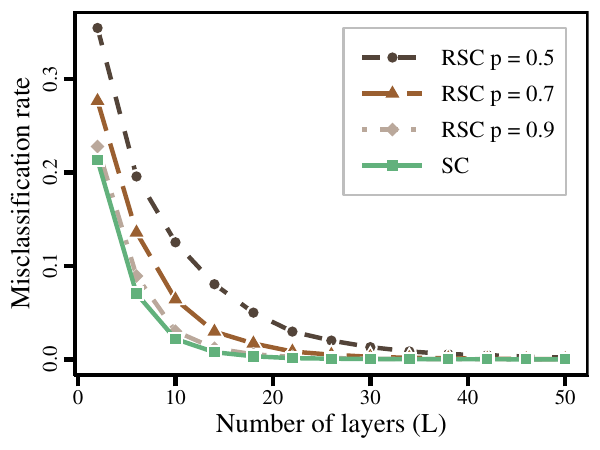}
        \caption{Effect of $L$, $q = 4$}
    \end{subfigure}
    \begin{subfigure}{0.325\textwidth}
        \includegraphics[width=\textwidth]{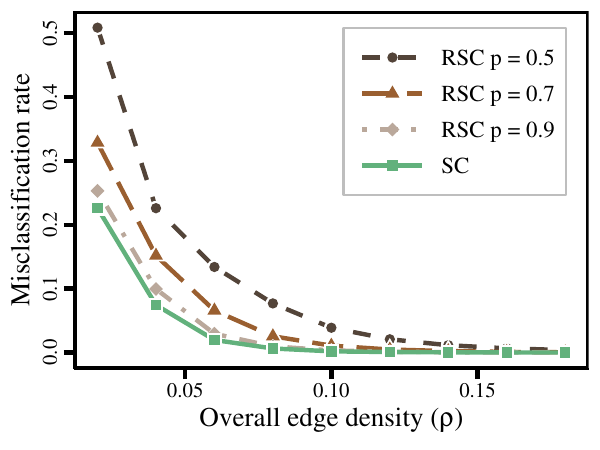}
        \caption{Effect of $\rho$, $q = 4$}
    \end{subfigure} \\
    \begin{subfigure}{0.325\textwidth}
    	\includegraphics[width=\textwidth]{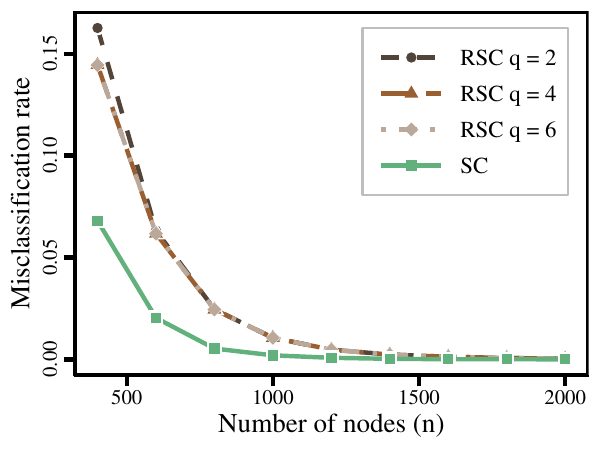}
        \caption{Effect of $n$, $p = 0.7$}
    \end{subfigure}
    \begin{subfigure}{0.325\textwidth}
        \includegraphics[width=\textwidth]{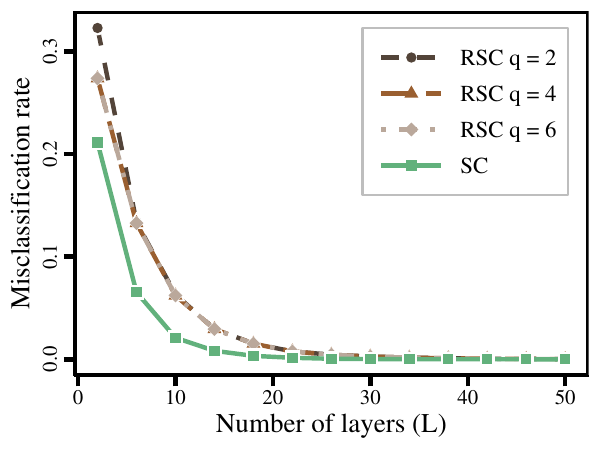}
        \caption{Effect of $L$, $p = 0.7$}
    \end{subfigure}
    \begin{subfigure}{0.325\textwidth}
        \includegraphics[width=\textwidth]{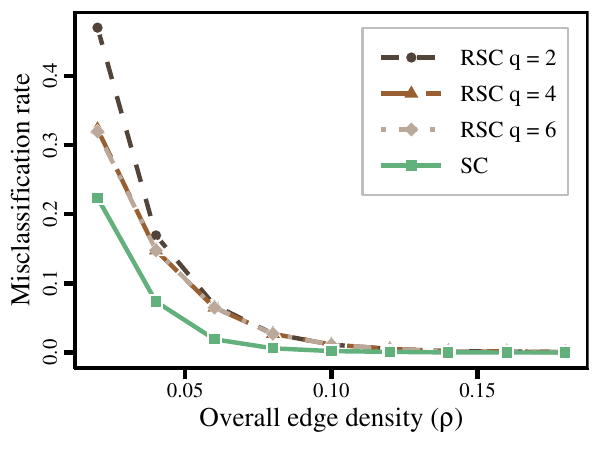}
        \caption{Effect of $\rho$, $p = 0.7$}
    \end{subfigure} 
    \caption{The misclassification rates of two methods under Model 2 with different parameters and hyperparameters.}
    \label{sim::model2}
\end{figure*}

\begin{figure*}[!htbp]
    \centering
    \begin{subfigure}{0.325\textwidth}
    	\includegraphics[width=\textwidth]{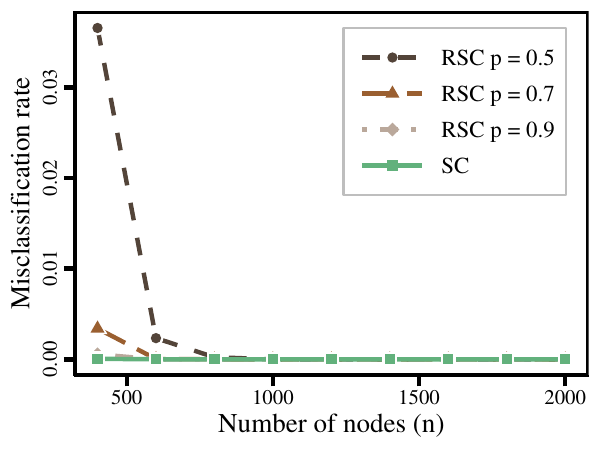}
    	\vspace{-0.13\textwidth}
        \caption{Effect of $n$, $q = 4$}
    \end{subfigure}
    \begin{subfigure}{0.325\textwidth}
        \includegraphics[width=\textwidth]{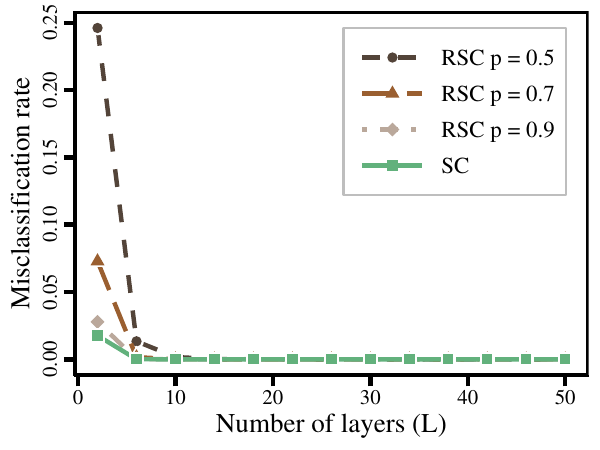}
    	\vspace{-0.13\textwidth}
        \caption{Effect of $L$, $q = 4$}
    \end{subfigure}
    \begin{subfigure}{0.325\textwidth}
        \includegraphics[width=\textwidth]{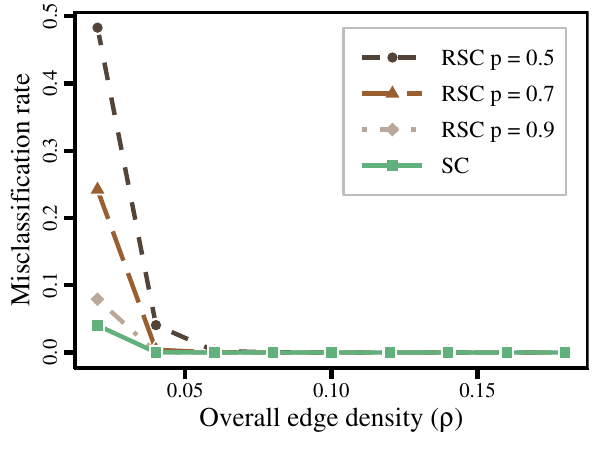}
    	\vspace{-0.13\textwidth}
        \caption{Effect of $\rho$, $q = 4$}
    \end{subfigure} \\
    \vspace{0.02\textwidth}
    \begin{subfigure}{0.325\textwidth}
    	\includegraphics[width=\textwidth]{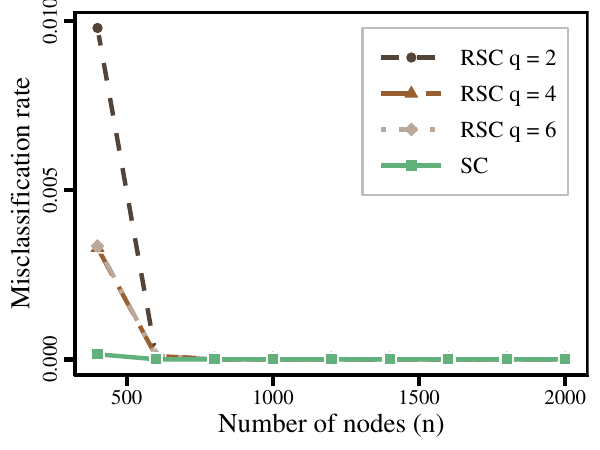}
    	\vspace{-0.13\textwidth}
        \caption{Effect of $n$, $p = 0.7$}
    \end{subfigure}
    \begin{subfigure}{0.325\textwidth}
        \includegraphics[width=\textwidth]{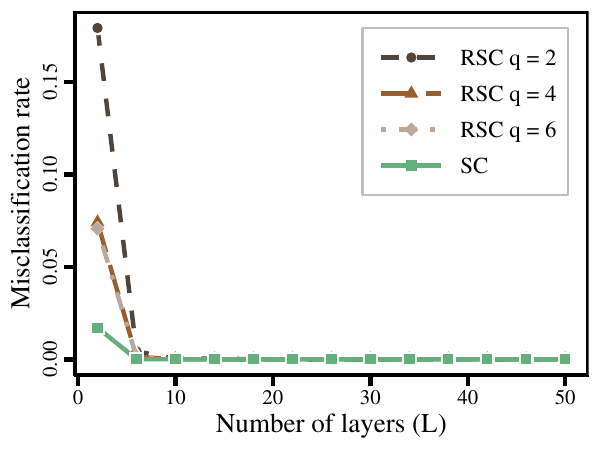}
    	\vspace{-0.13\textwidth}
        \caption{Effect of $L$, $p = 0.7$}
    \end{subfigure}
    \begin{subfigure}{0.325\textwidth}
        \includegraphics[width=\textwidth]{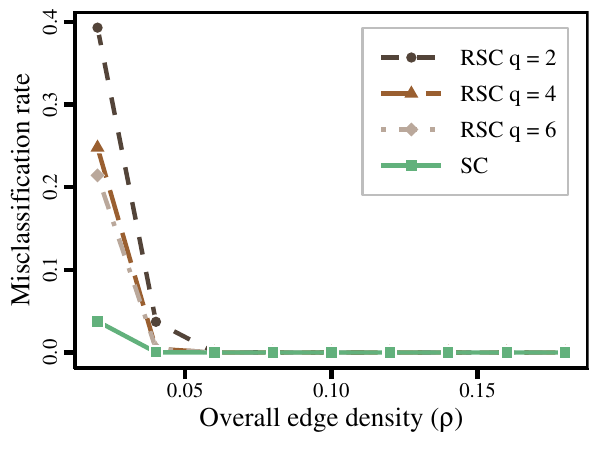}
    	\vspace{-0.13\textwidth}
        \caption{Effect of $\rho$, $p = 0.7$}
    \end{subfigure} 
    \caption{The misclassification rates of two methods under Model 3 with different parameters and hyperparameters.}
    \label{sim::model3}
\end{figure*}

\begin{figure*}[!htbp]
    \centering
    \begin{subfigure}{0.325\textwidth}
    	\includegraphics[width=\textwidth]{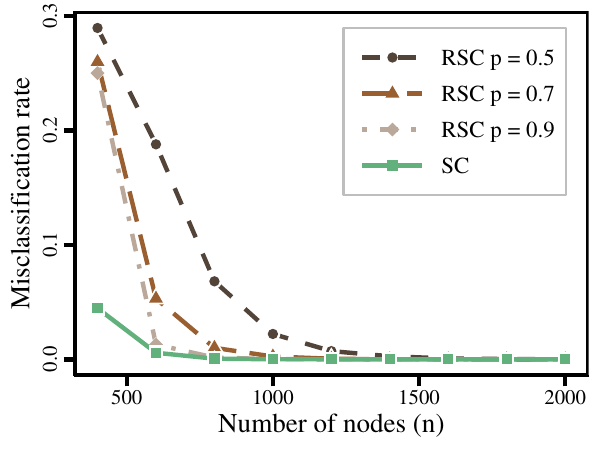}
    	\vspace{-0.13\textwidth}
        \caption{Effect of $n$, $q = 4$}
    \end{subfigure}
    \begin{subfigure}{0.325\textwidth}
        \includegraphics[width=\textwidth]{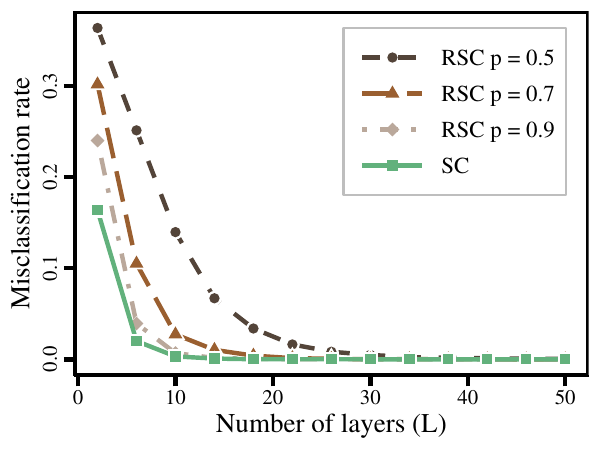}
    	\vspace{-0.13\textwidth}
        \caption{Effect of $L$, $q = 4$}
    \end{subfigure}
    \begin{subfigure}{0.325\textwidth}
        \includegraphics[width=\textwidth]{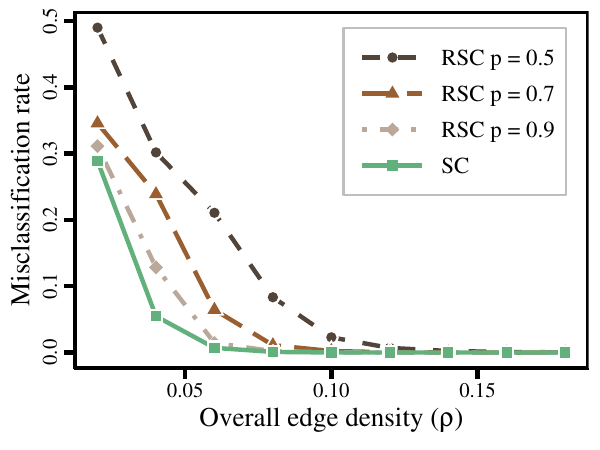}
    	\vspace{-0.13\textwidth}
        \caption{Effect of $\rho$, $q = 4$}
    \end{subfigure} \\
    \vspace{0.02\textwidth}
    \begin{subfigure}{0.325\textwidth}
    	\includegraphics[width=\textwidth]{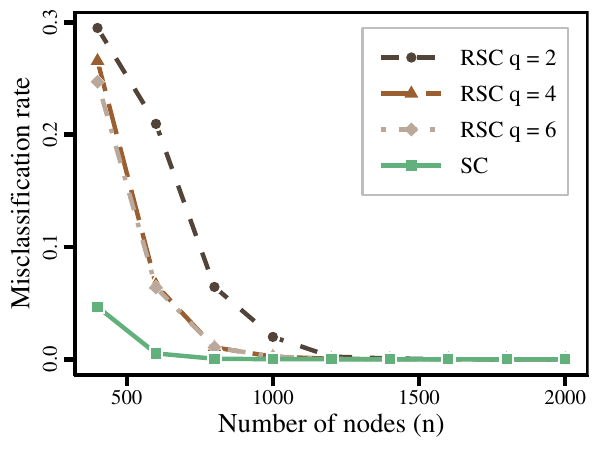}
    	\vspace{-0.13\textwidth}
        \caption{Effect of $n$, $p = 0.7$}
    \end{subfigure}
    \begin{subfigure}{0.325\textwidth}
        \includegraphics[width=\textwidth]{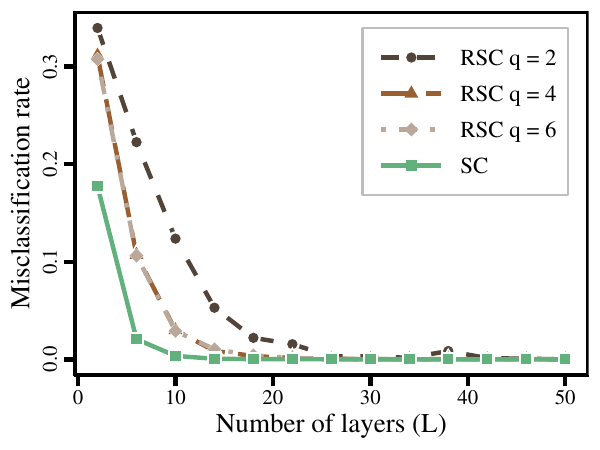}
    	\vspace{-0.13\textwidth}
        \caption{Effect of $L$, $p = 0.7$}
    \end{subfigure}
    \begin{subfigure}{0.325\textwidth}
        \includegraphics[width=\textwidth]{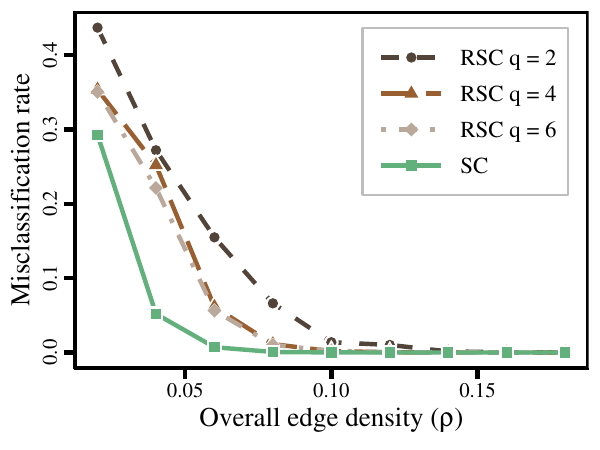}
    	\vspace{-0.13\textwidth}
        \caption{Effect of $\rho$, $p = 0.7$}
    \end{subfigure} \\
    \vspace{0.01\textwidth}       
    (I) row clustering \\
    \vspace{0.02\textwidth}       
    \begin{subfigure}{0.325\textwidth}
    	\includegraphics[width=\textwidth]{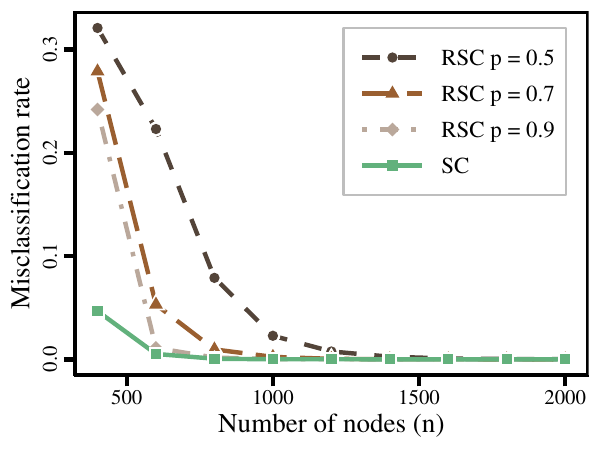}
    	\vspace{-0.13\textwidth}
        \caption{Effect of $n$, $q = 4$}
    \end{subfigure}
    \begin{subfigure}{0.325\textwidth}
        \includegraphics[width=\textwidth]{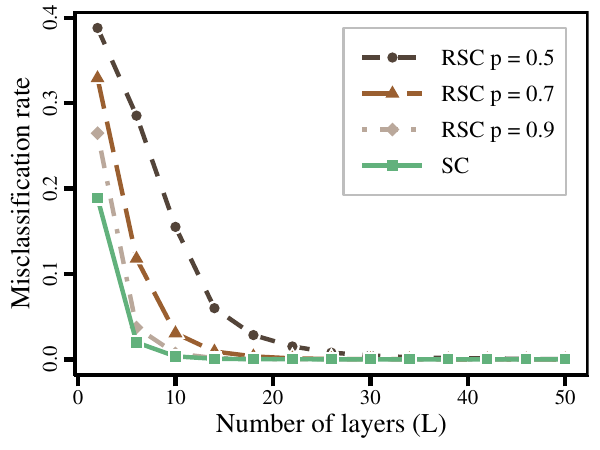}
    	\vspace{-0.13\textwidth}
        \caption{Effect of $L$, $q = 4$}
    \end{subfigure}
    \begin{subfigure}{0.325\textwidth}
        \includegraphics[width=\textwidth]{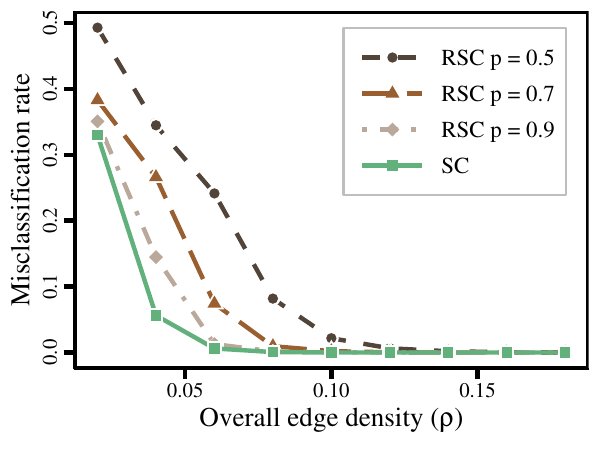}
    	\vspace{-0.13\textwidth}
        \caption{Effect of $\rho$, $q = 4$}
    \end{subfigure} \\
    \vspace{0.02\textwidth}
    \begin{subfigure}{0.325\textwidth}
    	\includegraphics[width=\textwidth]{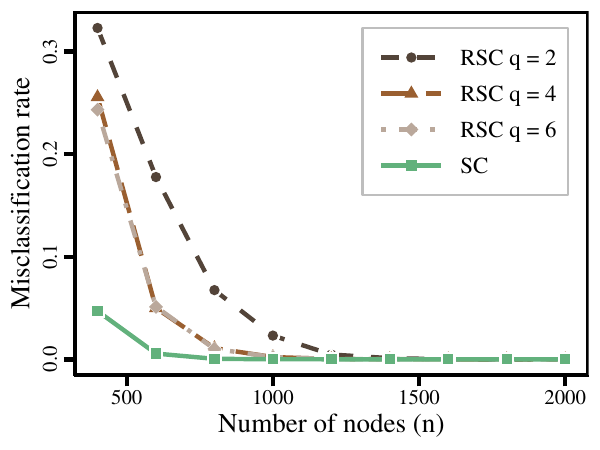}
    	\vspace{-0.13\textwidth}
        \caption{Effect of $n$, $p = 0.7$}
    \end{subfigure}
    \begin{subfigure}{0.325\textwidth}
        \includegraphics[width=\textwidth]{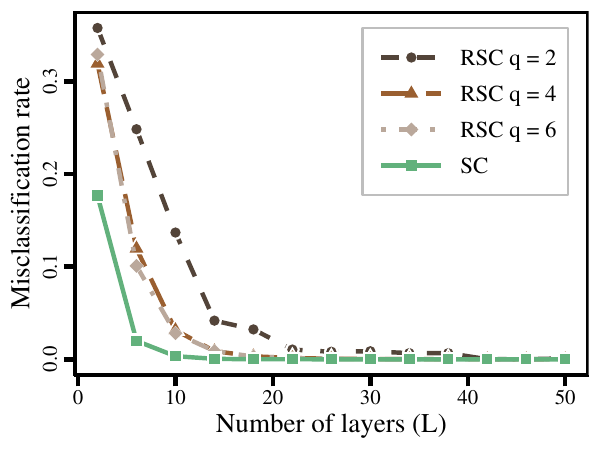}
    	\vspace{-0.13\textwidth}
        \caption{Effect of $L$, $p = 0.7$}
    \end{subfigure}
    \begin{subfigure}{0.325\textwidth}
        \includegraphics[width=\textwidth]{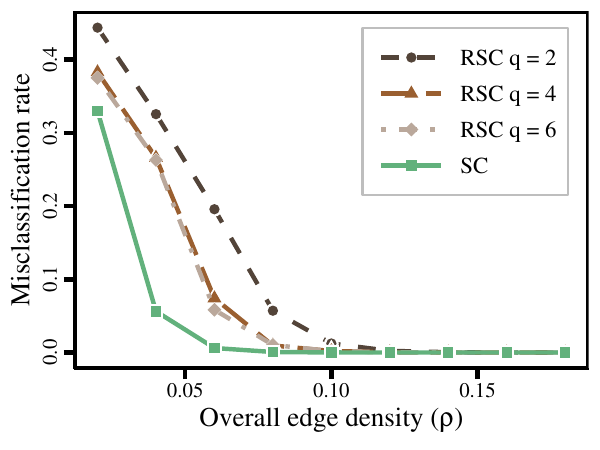}
    	\vspace{-0.13\textwidth}
        \caption{Effect of $\rho$, $p = 0.7$}
    \end{subfigure}\\
    \vspace{0.01\textwidth}       
    (II) column clustering \\
    \vspace{0.02\textwidth}   
    \caption{The misclassification rates of two methods under Model 4 with different parameters and hyperparameters.}
    \label{sim::model4}
\end{figure*}

The average results over 50 replications corresponding to Models 1 to 4 are shown in Figures \ref{sim::model1} to \ref{sim::model4}, respectively. It is evident that for all parameter settings, the proposed \texttt{RSC} performs closely to the performance of the \texttt{SC} as $n$, $L$ and $\rho$ increase, which is consistent with the theoretical results. Moreover, as expected, the performance of \texttt{RSC} improves with increasing the sampling rate and the power parameter. The above observation, along with the significantly reduced computational complexity achieved by \texttt{RSC}, demonstrates its effectiveness in handling large-scale multi-layer networks.

%Recall that the random projection in \texttt{RSC} is implemented using the randomized block Krylov method \citep{musco2015randomized}. In what follows, we compare \texttt{RSC} with its counterpart using the subspace-based random projection method \citep{rokhlin2010randomized, halko2011finding} employed in \citep{zhang2022randomized, guo2020randomized}, denoted by \texttt{RSC-S}. Specifically, in \texttt{RSC-S}, the line 9 of Algorithm \ref{rsc} is replaced with $G=\mybar{M}^{2q+1}\Omega$. The averaged results over 50 replications are displayed in Figure \ref{sim::wb}. We can observe that our method \texttt{RSC} outperforms the \texttt{RSC-S} method in all settings. \xiaoc{Let us discuss. I think the numerical performance of two methods are close. }
%\begin{figure*}[!htbp]
%    \centering
%    \begin{subfigure}{0.325\textwidth}
%    	\includegraphics[width=\textwidth]{rscs_n.pdf}
%    	\vspace{-0.13\textwidth}
%        \caption{Effect of $n$, $p = 0.7$}
%    \end{subfigure}
%    \begin{subfigure}{0.325\textwidth}
%        \includegraphics[width=\textwidth]{rscs_l.pdf}
%    	\vspace{-0.13\textwidth}
%        \caption{Effect of $L$, $p = 0.7$}
%    \end{subfigure}
%    \begin{subfigure}{0.325\textwidth}
%        \includegraphics[width=\textwidth]{rscs_rho.pdf}
%    	\vspace{-0.13\textwidth}
%        \caption{Effect of $\rho$, $q = 0.7$}
%    \end{subfigure} 
%    \caption{Comparison of \texttt{RSC} and \texttt{RSC-S} with varying power parameters. The underlying multi-layer SBM is the same as in Model 1.}
%    \label{sim::wb}
%\end{figure*}

\section{Real data analysis}\label{sec::real}
In this section, we evaluate the proposed algorithm \texttt{RSC} on seven real datasets in terms of the computational efficiency and clustering accuracy.  

The seven datasets consist of both directed and undirected multi-layer networks, with the number of nodes ranging from one thousand to three million. The detailed information is as follows.
\begin{enumerate}[label=(\roman*)] 
	\item European email network \citep{paranjape2017motifs}. A collection of emails exchanged among members of a European research institution, with layers representing different time periods.
	\item Yeast landscape network \citep{costanzo2010genetic}. The multilayer genetic interaction of the Saccharomyces Cerevisiae, a species of yeast.
	\item Home genetic network \citep{stark2006biogrid}. The multilayer genetic and protein interactions of the Homo Sapiens.
	\item MoscowAthletics 2013 social network \citep{omodei2015characterizing}. Various types of social relationships among Twitter users during 2013 World Championships in Athletics, with layers corresponding to retweet, mentions, and replies observed between.
	\item Cannes2013 social network \citep{omodei2015characterizing}. Similar to the previous network, this dataset focuses on social interactions during the 2013 Cannes Film Festival.
	\item YouTube social network \citep{mislove2007measurement}. A social network of YouTube users and their friendship connections, with layers representing different time periods. 
	\item Stack overflow network \citep{paranjape2017motifs}. A network of interactions on the stack exchange website Stack Overflow, with layers representing different types of interactions.
\end{enumerate}
The basic statistics of these real multi-layer networks are presented in Table \ref{networks}.

\begin{table}[ht]
\centering
\caption{Basis statistics of the seven real multi-layer networks, including the number of nodes, the number of layers, the number of edges and the directness.}
\begin{tabular}{@{} l *{4}{c} @{}} 
\toprule
Networks & Nodes & Layers & Edges & Directness  \\
\midrule
European email network & 986 & 6 & 56,440 & Directed \\
Yeast landscape network & 4,458 & 4 & 8,450,408 & Undirected \\
Home genetic network  & 18,222 & 7 & 170,899 & Directed \\
MoscowAthletics 2013 social network & 88,804 & 3 & 210,250 & Directed \\
Cannes 2013 social network & 438,537 & 3 & 991,854 & Directed \\
Youtube social network & 3,223,589 & 6 & 20,621,018 & Undirected \\
Stack overflow network & 2,601,977 & 3 & 37,619,132 & Undirected \\
\bottomrule
\label{networks}
\end{tabular}
\end{table}

To examine the computational efficiency of our method, we run \texttt{RSC} on all seven real multi-layer networks and report the computational times. Table \ref{time} presents the average computational time (seconds) of our method over 50 replications, where all computations are performed on a machine with an Apple M2 processor and 16GB of memory, using R version 4.2.0. For comparison, we include the computational time of the original spectral clustering algorithm (denoted by \texttt{SC}), namely, the non-randomized counterpart of \texttt{RSC}, on two small-scale multi-layer networks, the European email network and the Yeast landscape network. The number of clusters in Table \ref{time} is determined by the scree plot of the approximated eigenvalues of the aggregated matrix $M$ for undirected networks, and $M^R$ and $M^C$ for directed networks. When the number of nodes is four thousand, our \texttt{RSC} proved to be nearly thirty times faster than \texttt{SC}. As the number of nodes increased to ten thousand, \texttt{SC} failed due to memory constraints, while \texttt{RSC} continued to perform rapid clustering, efficiently handling networks with up to millions of nodes. This demonstrates the robust efficiency of our randomized algorithm across various network sizes and types.

\begin{table}[!htbp]
\centering
\caption{Average time (seconds) for clustering seven real multi-layer networks using \texttt{RSC} and \texttt{SC}. A dash `-' indicates that the clustering could not proceed due to memory overflow.}
\begin{tabular}{@{} l *{3}{c} @{}} 
\toprule
Networks & Clusters & Time of \texttt{RSC} & Time of \texttt{SC} \\
\midrule
European email network & 2, 2 & 0.092 & 0.112  \\
Yeast landscape network & 5 & 1.4 & 38 \\
Home genetic network  & 3, 2 & 0.363 & - \\
MoscowAthletics 2013 social network & 3, 2 & 0.886 & - \\
Cannes 2013 social network & 5, 3 & 7.425 & - \\
Youtube social network & 4 & 53.523 & - \\
Stack overflow network & 5 & 51.225 & - \\
\bottomrule
\label{time}
\end{tabular}
\end{table}

To evaluate the clustering accuracy of the proposed algorithm \texttt{RSC}, we compare \texttt{RSC} with \texttt{SC}. We use the Adjusted Rand Index (ARI) and Adjusted Mutual Information (AMI) to measure the similarity between the clusters obtained by two methods. A higher ARI or AMI indicates greater similarity, with both metrics reaching a value of 1 when the clusters are identical. The numbers of clusters for both methods are determined by the scree plot of the approximated eigenvalues (in absolute value) of the aggregated matrix $M$ for undirected networks, and $M^R$ and $M^C$ for directed networks. {The scree plot typically shows an eigen-gap between the eigenvalues corresponding to the target clusters and the subsequent smaller eigenvalues, which suggests the number of clusters.} We set the sampling rate $p = 0.7$ and the power parameter $q = 2$ in \texttt{RSC}. We only compare the two methods on two relatively smaller-scale networks, the European email network and the Yeast landscape network, which is because when the number of nodes reaches tens of thousands, \texttt{SC} becomes impractical due to the high memory demands required for directly computing the square of the adjacency matrix, which can lead to memory exhaustion. The average results over 50 replications are presented in Table \ref{accuracy}. It is evident from the results that the proposed algorithm performs similarly with the non-randomized counterpart in clustering, indicated by the large ARI and AMI. 

\begin{table}[ht]
\centering
\caption{The similarity of the clustering results between \texttt{RSC} and \texttt{SC} is measured by ARI and AMI. Larger ARI and AMI indicate more similarity. }
\begin{tabular}{@{} l *{2}{c} @{}} 
\toprule
Networks & ARI & AMI \\
\midrule
European email network (row) & 0.965 & 0.904  \\
European email network (column) & 0.974 & 0.924  \\
Yeast landscape network & 0.929 & 0.902  \\
\bottomrule
\label{accuracy}
\end{tabular}
\end{table}

\section{Conclusion}\label{sec::con}
In this paper, our goal was to develop an efficient algorithm for community detection of large-scale multi-layer networks. To this end, we proposed a randomized spectral clustering algorithm, where we utilized the  random sampling to reduce the time complexity of squaring the adjacency matrices and used the random projection to accelerate the eigen-decomposition of the sum of squared sparsified adjacency matrices. The algorithm reduces both the time complexity and space complexity. In theory, we studied the misclassification error rates of the proposed algorithm under the multi-layer SBMs. The results showed that under certain conditions, the randomization did not substantially deteriorate the clustering accuracy. Finally, we conducted extensive experiments to show the efficiency of proposed algorithm on multi-layer networks with millions of nodes. In the future, it would be interesting to generalize the methodology and theory to other community detection methods, say the tensor-based methods, for multi-layer networks.

\appendix
\renewcommand{\thesection}{\Alph{section}}
\renewcommand{\thesubsection}{\Alph{subsection}}
\section*{Appendix} \label{appendix}
Appendix \ref{appendix extension} extends the method to multi-layer ScBMs and presents the corresponding theoretical results. Appendix \ref{appendix technical} provides the technical theorem and lemmas that are needed to prove the misclassification rates. Appendix \ref{appendix proof} includes the proof of the theorem in the main text. Appendix \ref{appendix auxiliary} contains auxiliary lemmas.

\subsection{Extension to multi-layer ScBMs}\label{appendix extension}
We extend the \texttt{RSC} algorithm to the spectral co-clustering of multi-layer directed networks, which are assumed to follow the ScBMs. 

Compared to standard clustering, where only one set of clusters is obtained, co-clustering aims to obtain two sets of clusters (i.e., co-clusters), one for the row clusters containing nodes with similar sending patterns, and the other for the column clusters containing nodes with similar receiving patterns.

The multi-layer ScBM is designed to capture the co-clusters in multi-layer directed networks. Suppose there are $L$ layers and $n$ nodes. For each layer $l \in [L]$, the multi-layer ScBM assumes that the network adjacency matrix $A_l$ is independently generated as follows,
\begin{equation}\label{mscbm}
A_{l,ij}\sim_{i.i.d.}{\rm Bernoulli}(B_{l,g_i^y g_j^z})\; {\rm if} \; i\neq j \quad {\rm and} \quad A_{l,ij}=0\; {\rm if}\; i=j,
\end{equation}
where $g_i^y \in \{1,\ldots,K_y\}$ and $g_i^z \in \{1, \ldots, K_z\}$ denote the row and column community assignments of node $i$, for all $i \in [n]$. The matrix $B_l \in [0,1]^{K_y \times K_z}$ represents the block probability matrix for layer $l$. Here, $K_y$ and $K_z$, representing the numbers of row and column communities respectively, which can differ. The population of $A_l$ can be formulated as $P_l := Y B_l Z^T $, where $Y \in \{0, 1\}^{n \times K_y}$ and $Z \in \{0, 1\}^{n \times K_z}$ represent the row and column membership matrices, respectively.

As discussed in Section \ref{sec::prob}, to circumvent the potential clustering cancellation caused by directly summing the matrices $A_l$, utilizing the squared sum of adjacency matrices is more reasonable. We consider the spectral co-clustering method in \cite{su2024spectral} as the basis of our randomized algorithm. Specifically,  define
\begin{equation*}
\label{M_d}
  M^R:=\frac{1}{Ln}\sum_{l=1}^L (A_lA_l^T-D_l^{\rm out})\qquad {\rm and} \qquad M^C:=\frac{1}{Ln}\sum_{l=1}^L (A_l^TA^l-D_l^{\rm in}),
\end{equation*}
where $D_l^{\rm out}$ is a diagonal matrix consisting of the out-degree of $A_l$ with $D_{l,ii}^{\rm out}=\sum_j A_{l,ij}$ and $D_l^{\rm in}$ is a diagonal matrix consisting of the in-degree of $A_l$ with $D_{l,ii}^{\rm in}=\sum_j A_{l,ji}$. 
The spectral co-clustering obtains the row and column communities partition by applying $k$-means on the rows of the leading eigenvectors of $M^R$ and the leading eigenvectors of $M^C$, respectively.

Similar to the undirected multi-layer networks set-up, we utilize the randomized sketching techniques to accelerate the spectral co-clustering algorithm for directed multi-layer networks. 
The algorithm is similar to Algorithm \ref{rsc} with two main modifications: First, the random sampling step in \eqref{rs} is independently performed for each pair $1\leq i \neq j \leq n$. Second, both the bias-correction step in \eqref{dm} and the subsequent random-projection-based eigen-decomposition are performed on both row and column communities. Specifically, $\mybar{M}_l$ should be replaced with $\mybar{M}_l^R$ and $\mybar{M}_l^C$, where $\mybar{M}_l^R:= \frac{\tA_l\tA_l^T}{n}-\frac{\tilde{D}_{l}^{\rm out}}{np^2}$ and $\mybar{M}_l^C:= \frac{\tA_l^T\tA_l}{n}-\frac{\tilde{D}_{l}^{\rm in}}{np^2}$.
For simplicity, we will still refer to the randomized spectral co-clustering for multi-layer ScBMs as \texttt{RSC}.

Let $\widehat Y$ and $\widehat Z$ be the estimated row and column membership matrices obtained from the \texttt{RSC} algorithm for multi-layer ScBMs. Analogous to Theorem \ref{thm:mr}, we can also provide error bounds for the misclassification rates $\mathcal{L}(Y,\widehat{Y})$ and $\mathcal{L}(Z,\widehat{Z})$. We begin by presenting  some assumptions for multi-layer ScBMs, which are counterparts of Assumptions \ref{balcom} and \ref{rank} for multi-layer SBMs.

\renewcommand{\theassumption}{E1}
\begin{assumption}\label{E2}
For $k\in[K_y]\;(k\in[K_z])$, $n_k^y$ $(n_k^z)$ denotes the number of nodes within the $k$-th row $($column$)$ community.
The community sizes are balanced such that for any $k\in[K_y]$, $cn/K_y\leq n_k^y \leq Cn/K_y$, and for any $k\in[K_z]$, $cn/K_z\leq n_k^z \leq Cn/K_z$, where $c$ and $C$ are constants.
\end{assumption}

\renewcommand{\theassumption}{E2}
\begin{assumption}\label{E3}
$\sum_{l=1}^L B_{l,0}B_{l,0}^T$ has full rank $K_y$ with $\sigma_{\rm min}(\sum_{l=1}^L B_{l,0}B_{l,0}^T)\geq cL$ and $\sum_{l=1}^L B_{l,0}^TB_{l,0}$ has full rank $K_z$ with $\sigma_{\rm min}(\sum_{l=1}^L B_{l,0}^TB_{l,0})\geq cL$, where $\sigma_{\rm min}$ denotes the minimum eigenvalue of a matrix.
\end{assumption}

The following corollary provides upper bounds on the proportion of misclustered nodes for the multi-layer ScBM. The proof is similar to that in Theorem \ref{thm:mr}, and is therefore omitted.

\begin{corollary}
	Suppose Assumptions {\rm \ref{E2}} and {\rm \ref{E3}} hold, the following results hold respectively for two sparsity regimes.
\begin{enumerate}[label=(\roman*)] 
\item In regime \eqref{sparse-condition}, we have
\begin{align*}
	\max \{\mathcal{L}(Y,\widehat{Y}), \mathcal{L}(Z,\widehat{Z})\}\lesssim\max \{\frac{\log^{2}(L+n)}{p^{3}Ln^{2}\rho^2},\; \frac{1}{n^2}\}
\end{align*}
with probability larger than $1-O((L+n)^{- c_1}) - 2e^{-c_2K}$ for some positive constants $c_1$ and $c_2$.
\item In regime \eqref{dense-condition}, we have
\begin{align*}
	\max \{\mathcal{L}(Y,\widehat{Y}), \mathcal{L}(Z,\widehat{Z})\}\lesssim\max \{\frac{\log(L+n)}{pLn\rho},\;\frac{1}{n^2}\}
\end{align*}
with probability larger than $1-O((L+n)^{- c_3}) - 2e^{-c_4K}$ for some positive constants $c_3$ and $c_4$.
\end{enumerate}
\end{corollary}

\subsection{{Technical lemmas and theorem}}\label{appendix technical}
In this subsection, we provide the lemmas and theorem that will be used in the proof of the main theorem.
\begin{lemma}\label{sparse lemma}
	Let $A_l$ be the independent adjacency matrices generated by a multi-layer SBM satisfying the sparsity condition \eqref{sparse-condition} for all $1 \leq l \leq L$, and let $\tA_l$ represent the sparsified matrices with sampling probability $p$. Then, the following statements hold simultaneously with probability larger than $1-O((L+n)^{-c})$ for some positive constant $c$:
	\begin{enumerate}[label=(\roman*)] 
		\item $\max_{l, i} \tilde d_{l,i} \lesssim \log(L + n). $
		\item $\max_{i} \sum_{l=1}^L \tilde d_{l,i} \lesssim Ln\rho p. $
		\item $\|\sum_{l=1}^L\tA_l^2\| \lesssim Ln\rho / p.$
	\end{enumerate}

\end{lemma}
\emph{Proof.} 
%Recall the definition of $\tA_l$ in \eqref{rs}, we have
%\begin{equation*}
%	\ME |\tA_{l,ij}|^{\alpha} \leq |\frac{1}{p}|^{\alpha}pP_{l,ij} \leq \frac{\rho}{2p}\alpha!\left(\frac{1}{p}\right)^{\alpha-2}.
%\end{equation*}
%Thus, $\tA_{l,ij}$'s are $(\rho / p, 1/p)$-Bernstein. 
Recall the definition of $\tA_l$ in \eqref{rs}, where $\tA_{l,ij}$'s are independent and bounded. The first two statements are derived through a direct application of the standard Bernstein's inequality and the union bound.

The proof of the last part follows that of Lemma 8 in \cite{lei2020bias}, with the main difference being the tail conditions of the entries of $\tA_{l}$. We omit it here for brevity. \QEDA

\begin{lemma}\label{sparse e2}
	Under the same conditions as Lemma \ref{sparse lemma}, let $X_l:=\tA_l-\bar{P}_l$ be the noise matrices for all $1 \leq l \leq L$. Then $N_1$, the off-diagonal part of $\frac{1}{L}\sum_{l=1}^L X_l^2$, satisfies 
	\begin{equation*}
		\|N_{1}\|_2\lesssim \frac{n\rho\log(L+n)}{p^{3/2}L^{1/2}}
	\end{equation*}
	with probability at least $1-O((L+n)^{-c})$ for some constant $c>0$.	
\end{lemma}
\emph{Proof.} We use the decoupling techniques outlined in \cite{lei2020bias} and \cite{su2024spectral} to establish the proof. To use the decoupling argument, define 
	\begin{gather*}
		\widetilde{N}_1 = \frac{1}{L}\sum_{l=1}^L \sum_{(i,\;j)\neq(i',\;j')}X_{l,ij}X^{\rm ic}_{l,i'j'}e_ie_{i'}^TI_{\{j = j'\}}, \\ 
		\widetilde{N}_2 = \frac{1}{L}\sum_{l=1}^L \sum_{(i,\;j)}X_{l,ij} X^{\rm ic}_{l, ij}e_ie_i^T,
	\end{gather*}
	and 
	\begin{equation*}
		\widetilde{N} = \frac{1}{L}\sum_{l=1}^L X_lX^{\rm ic}_l = \widetilde{N}_1 + \widetilde{N}_2.
	\end{equation*}
	where $\widetilde{N}_1$ is the zero mean off-diagonal part and $\widetilde{N}_2$ is the diagonal part. Here, $e_i$ is the standard basis vector in $\mathbb{R}^n$ and pairs ${(i,j), \;(i',j')} \in \{1,2,\ldots,n\}^2$. $X^{\rm ic}_l$ is an independent copy of $X_l$ for all $1 \leq l \leq L$. Note that $\|\widetilde{N}_1\|_2 \leq \|\widetilde{N}\|_2 + \|\widetilde{N}_2\|_2$, we separately control the spectral norms of $\widetilde{N}$ and $\widetilde{N}_2$ .
	
	Recall that $X^{\rm ic}_l = \tA^{\rm ic}_l - \bar P_l$, where $\tA^{\rm ic}_l$ is an independent copy of $\tA_l$, we reformulate $\widetilde{N}$ as
	\begin{equation*}
		L\widetilde{N} = \sum_{l=1}^LX_lX^{\rm ic}_l = \sum_{l=1}^LX_l\tA^{\rm ic}_l - \sum_{l=1}^LX_l\bar P_l.
	\end{equation*}
	We first apply Lemma \ref{sparse lemma} and \ref{lei-bound1} to control $\|\sum_{l=1}^LX_l\tA^{\rm ic}_l\|_2$ conditioning on $\tA^{\rm ic}_1, \ldots, \tA^{\rm ic}_L$. Note that $\max_l\|\tA^{\rm ic}_l\|_{2,\infty} \leq \max_{l,i}\tilde d_{l,i} / p \lesssim \log(L + n)/p$ and  $n\|\sum_{l=1}^L\tA^{\rm ic}_l\tA^{\rm ic}_l\|_2 \lesssim Ln^2\rho/p$ with high probability, we have
	\begin{equation*}
		\left\|\sum_{l=1}^LX_l\tA^{\rm ic}_l\right\|_2 \leq cL^{1/2}n\rho\log(L+n)/p^2
	\end{equation*}
	with probability larger than $1-O((L+n)^{-c})$ for some constant $c>0$. Combining this with \eqref{e1}, it follows that with probability at least $1-O((L+n)^{-c})$, 
	\begin{equation*}
		\|\widetilde{N}\|_2 \lesssim \frac{n\rho\log(L+n)}{L^{1/2}p^{3/2}}.
	\end{equation*}
	
	Recall that $\widetilde{N}_2$ is a diagonal matrix whose $i$th diagonal element is $\frac{1}{L}\sum_{l=1}^L\sum_{j=1}^nX_{l,ij}X^{\rm ic}_{l,ij}$. Using standard Bernstein's inequality and the union bound, we have with probability at least $1-O((L+n)^{-c})$,
	\begin{equation*}
		\|\widetilde{N}_2\|_2 \lesssim \frac{n^{1/2}\rho\log^{1/2}(L + n)}{L^{1/2}p^{3/2}}.
	\end{equation*}
	
	The claim follows by combining the bounds on $\|\widetilde{N}\|_2$ and $\|\widetilde{N}_2\|_2$ together with the decoupling inequality (Theorem 1 in \cite{de1995decoupling}).  \QEDA

\begin{theorem}
\label{thm:rs}
The following results hold respectively for two sparsity regimes.
\begin{enumerate}[label=(\roman*)] 
\item In regime \eqref{sparse-condition}, we have
\begin{align*}
	\|\mybar{M}-\Pi\|_2\lesssim\max \{\frac{\rho\log(L+n)}{p^{3/2}L^{1/2}},\;\rho^2\}
\end{align*}
with probability larger than $1-O((L+n)^{-c_1})$ for some constant $c_1>0$.

\item In regime \eqref{dense-condition}, we have
\begin{align*}
	\|\mybar{M}-\Pi\|_2\lesssim\max \{\frac{n^{1/2}\rho^{3/2}\log^{1/2}(L+n)}{p^{1/2}L^{1/2}},\;\rho^2\}
\end{align*}
with probability larger than $1-O((L+n)^{-c_2})$ for some constant $c_2>0$.
\end{enumerate}
\end{theorem}
\emph{Proof.} We utilize the concentration tools developed in \citet{lei2020bias} to prove.
Recalling the decomposition outlined in Section \ref{bias adjustment}, we obtain the following decomposition,
\begin{equation*}%\phantomsection
	n\mybar{M}-n\Pi= \frac{1}{L}\sum_{l=1}^L (n\mybar{M}_l-n\Pi_l):=E_0+E_1+E_2+E_3- \frac{1}{L}\sum_{l=1}^L\frac{\tilde{D}_l}{p^2},
\end{equation*}
with $E_{i}:= \frac{1}{L}\sum_{l=1}^L E_{l,i}$, where $E_{l,i}$ is the same as those in (\ref{composion2}). We proceed to bound these error terms, respectively.

For $E_{0}:= \frac{1}{L}\sum_{l=1}^L\left({-P_l}\cdot{\rm diag}(P_l)-{\rm diag}(P_l)\cdot P_l +({\rm diag}(P_l))^2\right)$, it is easy to see that
\begin{align*}
\|E_{0}\|_2\leq 2\max_l\|{\rm diag}(P_l)\|_2\max_l \|P_l\|_{\rm F}+\max _l\|{\rm diag}(P_l)\|_2^2\leq Cn\rho^2.
\end{align*}

For $E_{1}:= \frac{1}{L}\sum_{l=1}^L \left(X_l (P_l-{\rm diag}(P_l))+ (P_l-{\rm diag}(P_l))X_l\right)$, we will make use of the concentration inequality of \citet{lei2020bias} (see also Lemma \ref{lei-bound1}). For that, we now show that each entry of $pX_l$ satisfies the Bernstein condition (see Definition \ref{bernstein}). With a slight abuse of notation, for $i\neq j$, we define $F_l:=pX_{l,ij}=p(\tA_{l,ij}-P_{l,ij})$. Specifically,
\begin{align*}
	F_l=\begin{cases}
	1-pP_{l,ij}\;, &{\rm with\; probability}\, \; pP_{l,ij},\nonumber\\
	-pP_{l,ij}\;, &{\rm with\; probability} \,\; 1-pP_{l,ij},\nonumber
	\end{cases}
\end{align*}
and for $\alpha\geq 2$, we have
\begin{align}\label{bern}
	\ME |F_l^{\alpha}|&\leq |1-pP_{l,ij}|^{\alpha}pP_{l,ij}+|pP_{l,ij}|^{\alpha} (1-pP_{l,ij})\nonumber\\
	&\leq (1-pP_{l,ij})pP_{l,ij}+pP_{l,ij}(1-pP_{l,ij})\nonumber\\
	&\leq 2pP_{l,ij}\leq \frac{v_1}{2}\alpha!R_1^{\alpha-2},
\end{align}
where $v_1:=4p\rho$ and $R_1:=1$. Thus, $pX_{l,ij}$'s are $(v_1,R_1)$-Bernstein. We also have $\|\sum_{l=1}^L(P_l-{\rm diag}(P_l))^2\|_2\leq \sum_{l=1}^L\|P_l-{\rm diag}(P_l)\|_{\rm F}^2\leq L n^2\rho^2$ and $\max_l\|P_l-{\rm diag}(P_l)\|_{2,\infty}\leq \sqrt{n}\rho$. By using Lemma \ref{lei-bound1}, we have
\begin{align}\label{e1bound}
	\MP(pL\|E_{1}\|_2\geq t)\leq 4n\cdot \exp\left\{\frac {- t^2/2}{4pLn^3\rho^3+\sqrt{n}\rho t}\right\}.
\end{align}
Choosing $t=p^{1/2}\rho^{3/2}n^{3/2}L^{1/2} \log^{1/2} (L+n)$,  it follows that
$$\sqrt{n}\rho t \lesssim pL n^{3}\rho^3,$$
which requires $\rho \gtrsim \log^{1/2}(L+n)/(pn^2L)$ and the latter is automatically met for both of our regimes \eqref{sparse-condition} and \eqref{dense-condition}. As a result, \eqref{e1bound} implies that
\begin{align}\label{e1}
	\|E_{1}\|_2\lesssim \frac{n^{3/2}\rho^{3/2}\log^{1/2}(L+n)}{p^{1/2}L^{1/2}}
\end{align}
with probability larger than $1-O((L+n)^{-c})$ for some constant $c>0$.

For $E_{2}:= \frac{1}{L}\sum_{l=1}^L\left(X_l^2-{\rm diag}(X^2_l)\right)$, we discuss the upper bounds respectively under the sparse regime \eqref{sparse-condition} and dense regime \eqref{dense-condition}.

Under the sparse regime \eqref{sparse-condition}, Lemma \ref{sparse e2} implies that 	\begin{equation}\label{e2sparse}
		\|E_2\|_2\lesssim \frac{n\rho\log(L+n)}{p^{3/2}L^{1/2}}
	\end{equation}
	with probability at least $1-O((L+n)^{-c})$ for some constant $c>0$.	
	
Under the dense regime \eqref{dense-condition}, we will make use of Lemma \ref{lei-bound2}. Given an independent copy $pX^{\rm ic}_l$ of $pX_l$, and similar to the derivation of \eqref{bern}, we can show that $p^2X_{l,ij}X^{\rm ic}_{l,ij}$ is $(v'_2,R'_2)$-Bernstein with $v'_2=8p^2\rho^2$ and $R'_2=1$. Then by Lemma \ref{lei-bound2}, we obtain with probability larger than $1-O((L+n)^{-c})$ that
\begin{align*}\label{e2}
	\|Lp^2E_{2}\|_2& \lesssim p\rho n \log(L+n)\sqrt{L}+\sqrt{p\rho}\sqrt{Ln}\log^{3/2}(L+n)+p\rho \sqrt{Ln}\log(L+n)+\log^2(L+n).
\end{align*}
Under \eqref{dense-condition}, we have with probability at least $1-O((L+n)^{-c})$ for some constant $c > 0$,
\begin{equation}\label{e2dense}
	\|E_2\|_2\lesssim \frac{n\rho\log(L+n)}{\sqrt{L}p}.
\end{equation}

For $E_3- \frac{1}{L}\sum_{l=1}^L\frac{\tilde{D}_l}{p^2}$, we have derived in \eqref{e3com} that
\begin{align*}
	\|E_3-\frac{1}{L}\sum_{l}\frac{\tilde{D}_l}{p^2}\|_2\leq Cn\rho^2.
\end{align*}

Finally, we need to compare the three terms $E_1$, $E_2$ and $E_3$ ($E_0$ is of the same order with $E_3$). The following results can be observed. 
\begin{enumerate}[label=(\roman*)] 
\item Under the regime (\ref{sparse-condition}), the bound for $E_1$ is smaller than the bound \eqref{e2sparse} for $E_2$ and thus 
	$$\|\mybar{M}-\Pi\|_2\lesssim\max \{\frac{\rho\log(L+n)}{p^{3/2}L^{1/2}},\;\rho^2\}.$$
\item Under the regime (\ref{dense-condition}), the bound \eqref{e2dense} for $E_2$ is smaller than that for $E_1$ and thus 
	$$\|\mybar{M}-\Pi\|_2\lesssim\max \{\frac{n^{1/2}\rho^{3/2}\log^{1/2}(L+n)}{p^{1/2}L^{1/2}},\;\rho^2\}.$$
\end{enumerate}
\QEDA

\subsection{Proof of main theorem}\label{appendix proof}
\subsubsection*{Proof of Theorem \ref{thm:mr}}
Consider that the approximated eigenspace $\hat U$ used for $k$-means clustering in Algorithm \ref{rsc} is actually the eigenvector of $\hat M := QQ^T\mybar{M} QQ^T$.  We first use the concentration bound of $\mybar{M}$  around the population matrix $\Pi$ and the argument about the low-rank randomized approximation in \cite{musco2015randomized} to bound the deviation of $\hat M$ from $\Pi$, recalling the definition of $\Pi$ in \eqref{Pi}. We begin by noting that
\begin{align}
	\|\hat M - \Pi\|_2 &= \|QQ^T\mybar{M} QQ^T - \Pi\|_2 \nonumber \\
	& \leq \|\mybar{M} - \Pi\|_2 + \|QQ^T\mybar{M} QQ^T - \mybar{M}\|_2 \nonumber \\
	& \leq \|\mybar{M} - \Pi\|_2 + \|QQ^T\mybar{M} (QQ^T - \mybar{M})\|_2 + \|QQ^T\mybar{M} - \mybar{M}\|_2 \nonumber \\
	& \leq \|\mybar{M} - \Pi\|_2 + 2 \|\mybar{M} -QQ^T\mybar{M}\|_2.
\end{align}

For the second term $\|\mybar{M} -QQ^T\mybar{M}\|_2$, by the Theorem 1 of \cite{musco2015randomized}, when $q \asymp \log n /\sqrt{\varepsilon}$, the following inequality holds with probability at least $1 - 2e^{-cK}$,
\begin{equation*}
	\|\mybar{M} -QQ^T\mybar{M}\|_2 \leq (1 + \varepsilon)\|\mybar{M} - \mybar{M}_K\|_2,
\end{equation*}
where $\mybar{M}_K$ is $\mybar{M}$ projected onto the space spanned by its top $K$ singular vectors, $\epsilon \in (0, 1)$ is the error, and $c$ is some positive constant. Note that $\|\mybar{M} - \mybar{M}_K\|_2 = \sigma_{K + 1}(\mybar{M})$ and $\Pi$ is of rank $K$, then we have
\begin{equation*}
	\|\mybar{M} -QQ^T\mybar{M}\|_2 \leq  (1 + \varepsilon)\sigma_{K + 1}(\mybar{M}) = (1 + \varepsilon)\left(\sigma_{K + 1}(\mybar{M}) - \sigma_{K + 1}(\Pi)\right) \leq (1 + \varepsilon) \|\mybar{M} - \Pi\|_2,
\end{equation*}
where $\sigma_{K+1}(\cdot)$ denotes the $K+1$th largest eigenvalue of a symmetric matrix. 

As a result, we have with high probability
\begin{equation*}
	\|\hat M - \Pi\|_2 \leq (3+2\varepsilon)\|\mybar{M} - \Pi\|_2.
\end{equation*}
The upper bound of $\|\mybar{M} - \Pi\|_2$ is provided in Theorem \ref{thm:rs}.

Then, by the Davis-Kahan sin$\Theta$ theorem (Lemma \ref{lem:DK}), there exists a $K \times K$ orthogonal matrix $O$ such that
\begin{equation*}
	\|\hat{U}-UO\|_{\rm F} \leq\frac{2\sqrt{2K}\|\mybar{M}-\Pi\|_2}{\sigma_K(\Pi)} \leq \frac{2\sqrt{2K}\|\mybar{M}-\Pi\|_2}{cn\rho^2}.
\end{equation*}
Here, the last inequality comes from the given assumptions that the community sizes are balanced and the minimum eigenvalue of $\sum_{l=1}^L B_{l,0}^2$ is at least $cL$, ensuring that $\sigma_K(\Pi) \geq cn\rho^2$ for some constant $c > 0$. Finally, by using Lemmas \ref{kmeans} and \ref{aSBM-eigen}, we are able to obtain the desired bound for the misclassification rate. \QEDA

\subsection{Auxiliary lemmas} \label{appendix auxiliary}
This subsection includes the auxiliary lemmas used to prove the theorems in the paper.

\begin{definition}[Bernstein condition]
\label{bernstein}
A zero-mean random variable $X$ with ${\rm Var}(X) = v$ is said to be $(v,R)$-Bernstein if there exists a constant $R$ such that for all $\alpha\geq 2$, the $\alpha$-th moment satisfies $\ME[|X|^{\alpha}]\leq \frac{1}{2}v\alpha!R^{\alpha-2}$.
\end{definition}

\begin{lemma}[Theorem 3 in \citet{lei2020bias}]
\label{lei-bound1}
 Let $X_1,...,X_L$ be a sequence of independent $n\times r$ random matrices with zero-mean independent entries being $(v_1, R_1)$-Bernstein. Let $H_1,...H_L$ be any sequence of $r\times n$ non-random matrices. Then for all $t>0$,
$$\MP(\|\sum_{l=1}^L X_lH_l\|_2\geq t)\leq 4n\exp\left(-\frac{t^2/2}{v_1n\|\sum_{l=1}^L H_l^T H_l\|_2 + R_1 \max_l\|H_l\|_{2,\infty} t}\right).$$
\end{lemma}

\begin{lemma}[Theorem 4 in \citet{lei2020bias}]
\label{lei-bound2}
Let $X_1,X_2,...,X_L$ be independent $n\times n$ random symmetric matrices with independent diagonal and upper diagonal entries. For any $l\in[L]$ and any pair $(i, j)$ where $i \leq j$, $X_{l,ij}$ is $(v_1, R_1)$-Bernstein, and $X_{l,ij}\tilde{X}_{l,ij}$ is $(v'_2, R'_2)$-Bernstein where $\tilde{X}_{l,ij}$ is an independent copy of $X_{l,ij}$. Then with probability larger than $1-O((n+L)^{-1})$,
\begin{align*}
\|\sum_{l=1}^L X_l^2-{\rm diag}(\sum_{l=1}^LX_l^2) \|_2\leq C&\Big[v_1 n \log (L+n)\sqrt{L} +\sqrt{v_1}R_1\sqrt{Ln}\log^{3/2}(L+n)\nonumber\\
&\quad\quad+\sqrt{v'_2} \sqrt{Ln}\log (L+n)+(R_1^2+R'_2)\log^2(L+n) \Big].
\end{align*}
Moreover, if the squared entry $X_{l,ij}^2$ is $(v_2, R_2)$-Bernstein, then with probability larger than $1-O((n+L)^{-1})$,
\begin{align*}
\|{\rm diag}(\sum_{l=1}^L X_l^2)-\mathbb E({\rm diag}(\sum_{l=1}^L X_l^2)) \|_2\leq C\left[\sqrt{v_2}\log(L+n)\sqrt{Ln}+R_2\log(L+n)\right].
\end{align*}
\end{lemma}

\begin{lemma}[Davis-Kahan theorem]
	\label{lem:DK}
	Let $Q\in\mathbb R^{n\times n}$ be a rank $K$ symmetric matrix with the smallest nonzero singular value $\gamma_n$. Let $M$ be any symmetric matrix and $\hat{U},U\in \mathbb R^{n\times K}$ be the $K$ leading eigenvectors of $M$ and $Q$, respectively. Then there exists a $K\times K$ orthogonal matrix $O$ such that
	\[
	 \|\hat{U}-UO\|_{\rm F} \leq\frac{2\sqrt{2K}\|M-Q\|_2}{\gamma_n}.
	\]
\end{lemma}
\emph{Proof.} See \citet{lei2015consistency,chen2021spectral}, among others.
\QEDA

\begin{lemma}[Lemma 5.3 in \cite{lei2015consistency}]\label{kmeans}
	Let $U$ be an $n\times d$ matrix with $K$ distinct rows with minimum pairwise Euclidean norm separation $\gamma>0$. Let $\widehat{U}$ be another $n\times d$ matrix and $\widehat{\Theta}$ be a solution to $k$-means problem with input $\widehat{U}$, then the number of errors in $\widehat{\Theta}$ as an estimate of the row clusters of $U$ is no larger than $c\|\widehat{U} - U\|_{\rm F}^2\gamma^{-2}$ for some constant $c>0$.
\end{lemma}

\begin{lemma}
\label{aSBM-eigen}
Define the population matrix $Q=\Theta \bar{B}\Theta^\intercal$. Suppose $\bar{B}$ is of full rank and denote the eigen-decomposition of $Q$ by ${V}_{n\times K}{\Sigma}_{K\times K}{V}^\intercal_{K\times n}$. Then, $V=\Theta X$ for some given matrix $X$. Specifically, for $\Theta_{i\ast}=\Theta_{j\ast}$, we have ${V}_{i\ast}={V}_{j\ast}$; while for $\Theta_{i\ast}\neq\Theta_{j\ast}$, we have $\|{V}_{i\ast}-{V}_{j\ast}\|_2=\sqrt{(n_{g_i})^{-1}+(n_{g_j})^{-1}}$.
\end{lemma}
\emph{Proof.} See \citet{lei2015consistency,guo2020randomized,zhang2022randomized}, among others.

\QEDA

\bibliographystyle{chicago}
\bibliography{randmultinet}

\begin{thebibliography}{}

\bibitem[Ahfock et~al., 2021]{ahfock2021statistical}
Ahfock, D.~C., Astle, W.~J., and Richardson, S. (2021).
\newblock Statistical properties of sketching algorithms.
\newblock {\em Biometrika}, 108(2):283--297.

\bibitem[Arroyo et~al., 2021]{arroyo2021inference}
Arroyo, J., Athreya, A., Cape, J., Chen, G., Priebe, C.~E., and Vogelstein,
  J.~T. (2021).
\newblock Inference for multiple heterogeneous networks with a common invariant
  subspace.
\newblock {\em Journal of Machine Learning Research}, 22(142):1--49.

\bibitem[Baglama and Reichel, 2005]{baglama2005augmented}
Baglama, J. and Reichel, L. (2005).
\newblock Augmented implicitly restarted lanczos bidiagonalization methods.
\newblock {\em SIAM Journal on Scientific Computing}, 27(1):19--42.

\bibitem[Barbillon et~al., 2017]{barbillon2017stochastic}
Barbillon, P., Donnet, S., Lazega, E., and Bar-Hen, A. (2017).
\newblock Stochastic block models for multiplex networks: an application to a
  multilevel network of researchers.
\newblock {\em Journal of the Royal Statistical Society Series A: Statistics in
  Society}, 180(1):295--314.

\bibitem[Bhattacharyya and Chatterjee, 2018]{bhattacharyya2018spectral}
Bhattacharyya, S. and Chatterjee, S. (2018).
\newblock Spectral clustering for multiple sparse networks: I.
\newblock {\em arXiv preprint arXiv:1805.10594}.

\bibitem[Bianconi, 2018]{bianconi2018multilayer}
Bianconi, G. (2018).
\newblock {\em Multilayer networks: structure and function}.
\newblock Oxford university press.

\bibitem[Boccaletti et~al., 2014]{boccaletti2014structure}
Boccaletti, S., Bianconi, G., Criado, R., Del~Genio, C.~I., G{\'o}mez-Gardenes,
  J., Romance, M., Sendina-Nadal, I., Wang, Z., and Zanin, M. (2014).
\newblock The structure and dynamics of multilayer networks.
\newblock {\em Physics Reports}, 544(1):1--122.

\bibitem[Chen et~al., 2021]{chen2021spectral}
Chen, Y., Chi, Y., Fan, J., and Ma, C. (2021).
\newblock Spectral methods for data science: A statistical perspective.
\newblock {\em Foundations and Trends{\textregistered} in Machine Learning},
  14(5):566--806.

\bibitem[Costanzo et~al., 2010]{costanzo2010genetic}
Costanzo, M., Baryshnikova, A., Bellay, J., Kim, Y., Spear, E.~D., Sevier,
  C.~S., Ding, H., Koh, J.~L., Toufighi, K., Mostafavi, S., et~al. (2010).
\newblock The genetic landscape of a cell.
\newblock {\em Science}, 327(5964):425--431.

\bibitem[de~la Pe{\~n}a and Montgomery-Smith, 1995]{de1995decoupling}
de~la Pe{\~n}a, V.~H. and Montgomery-Smith, S.~J. (1995).
\newblock Decoupling inequalities for the tail probabilities of multivariate
  u-statistics.
\newblock {\em The Annals of Probability}, 23(2):806--816.

\bibitem[Deng et~al., 2024]{deng2024subsampling}
Deng, J., Huang, D., Ding, Y., Zhu, Y., Jing, B., and Zhang, B. (2024).
\newblock Subsampling spectral clustering for stochastic block models in
  large-scale networks.
\newblock {\em Computational Statistics \& Data Analysis}, 189:107835.

\bibitem[Drineas et~al., 2012]{drineas2012fast}
Drineas, P., Magdon-Ismail, M., Mahoney, M.~W., and Woodruff, D.~P. (2012).
\newblock Fast approximation of matrix coherence and statistical leverage.
\newblock {\em Journal of Machine Learning Research}, 13(1):3475--3506.

\bibitem[Guo et~al., 2023]{guo2020randomized}
Guo, X., Qiu, Y., Zhang, H., and Chang, X. (2023).
\newblock Randomized spectral co-clustering for large-scale directed networks.
\newblock {\em Journal of Machine Learning Research}, 24(380):1--68.

\bibitem[Halko et~al., 2011]{halko2011finding}
Halko, N., Martinsson, P.-G., and Tropp, J.~A. (2011).
\newblock Finding structure with randomness: Probabilistic algorithms for
  constructing approximate matrix decompositions.
\newblock {\em SIAM Review}, 53(2):217--288.

\bibitem[Han et~al., 2015]{han2015consistent}
Han, Q., Xu, K., and Airoldi, E. (2015).
\newblock Consistent estimation of dynamic and multi-layer block models.
\newblock In {\em International Conference on Machine Learning}, pages
  1511--1520. PMLR.

\bibitem[Jing et~al., 2021]{jing2021community}
Jing, B.-Y., Li, T., Lyu, Z., and Xia, D. (2021).
\newblock Community detection on mixture multilayer networks via regularized
  tensor decomposition.
\newblock {\em The Annals of Statistics}, 49(6):3181--3205.

\bibitem[Kivel{\"a} et~al., 2014]{kivela2014multilayer}
Kivel{\"a}, M., Arenas, A., Barthelemy, M., Gleeson, J.~P., Moreno, Y., and
  Porter, M.~A. (2014).
\newblock Multilayer networks.
\newblock {\em Journal of Complex Networks}, 2(3):203--271.

\bibitem[Lei et~al., 2020]{lei2020consistent}
Lei, J., Chen, K., and Lynch, B. (2020).
\newblock Consistent community detection in multi-layer network data.
\newblock {\em Biometrika}, 107(1):61--73.

\bibitem[Lei and Lin, 2023]{lei2020bias}
Lei, J. and Lin, K.~Z. (2023).
\newblock Bias-adjusted spectral clustering in multi-layer stochastic block
  models.
\newblock {\em Journal of the American Statistical Association},
  118(544):2433--2445.

\bibitem[Lei and Rinaldo, 2015]{lei2015consistency}
Lei, J. and Rinaldo, A. (2015).
\newblock Consistency of spectral clustering in stochastic block models.
\newblock {\em The Annals of Statistics}, 43(1):215--237.

\bibitem[Lei et~al., 2023]{lei2023computational}
Lei, J., Zhang, A.~R., and Zhu, Z. (2023).
\newblock Computational and statistical thresholds in multi-layer stochastic
  block models.
\newblock {\em arXiv preprint arXiv:2311.07773}.

\bibitem[Li and Zhu, 2021]{li2021randomized}
Li, H. and Zhu, Y. (2021).
\newblock Randomized block krylov subspace methods for trace and
  log-determinant estimators.
\newblock {\em BIT Numerical Mathematics}, 61:911--939.

\bibitem[Li et~al., 2020]{li2020network}
Li, T., Levina, E., and Zhu, J. (2020).
\newblock Network cross-validation by edge sampling.
\newblock {\em Biometrika}, 107(2):257--276.

\bibitem[Liu et~al., 2018]{liu2018global}
Liu, F., Choi, D., Xie, L., and Roeder, K. (2018).
\newblock Global spectral clustering in dynamic networks.
\newblock {\em Proceedings of the National Academy of Sciences},
  115(5):927--932.

\bibitem[Mahoney and Drineas, 2009]{mahoney2009cur}
Mahoney, M.~W. and Drineas, P. (2009).
\newblock Cur matrix decompositions for improved data analysis.
\newblock {\em Proceedings of the National Academy of Sciences},
  106(3):697--702.

\bibitem[Martinsson and Tropp, 2020]{martinsson2020randomized}
Martinsson, P.-G. and Tropp, J.~A. (2020).
\newblock Randomized numerical linear algebra: Foundations and algorithms.
\newblock {\em Acta Numerica}, 29:403--572.

\bibitem[Matias and Miele, 2017]{matias2017statistical}
Matias, C. and Miele, V. (2017).
\newblock Statistical clustering of temporal networks through a dynamic
  stochastic block model.
\newblock {\em Journal of the Royal Statistical Society Series B: Statistical
  Methodology}, 79(4):1119--1141.

\bibitem[Mislove et~al., 2007]{mislove2007measurement}
Mislove, A., Marcon, M., Gummadi, K.~P., Druschel, P., and Bhattacharjee, B.
  (2007).
\newblock Measurement and analysis of online social networks.
\newblock In {\em Proceedings of the 7th ACM SIGCOMM Conference on Internet
  Measurement}, pages 29--42.

\bibitem[Musco and Musco, 2015]{musco2015randomized}
Musco, C. and Musco, C. (2015).
\newblock Randomized block krylov methods for stronger and faster approximate
  singular value decomposition.
\newblock {\em Advances in Neural Information Processing Systems}, 28.

\bibitem[Noroozi and Pensky, 2024]{noroozi2022sparse}
Noroozi, M. and Pensky, M. (2024).
\newblock Sparse subspace clustering in diverse multiplex network model.
\newblock {\em Journal of Multivariate Analysis}, 203:105333.

\bibitem[Omodei et~al., 2015]{omodei2015characterizing}
Omodei, E., De~Domenico, M., and Arenas, A. (2015).
\newblock Characterizing interactions in online social networks during
  exceptional events.
\newblock {\em Frontiers in Physics}, 3:59.

\bibitem[Paranjape et~al., 2017]{paranjape2017motifs}
Paranjape, A., Benson, A.~R., and Leskovec, J. (2017).
\newblock Motifs in temporal networks.
\newblock In {\em Proceedings of the tenth ACM International Conference on Web
  Search and Data Mining}, pages 601--610.

\bibitem[Paul and Chen, 2016]{paul2016consistent}
Paul, S. and Chen, Y. (2016).
\newblock Consistent community detection in multi-relational data through
  restricted multi-layer stochastic blockmodel.
\newblock {\em Electronic Journal of Statistics}, 10(2):3807--3870.

\bibitem[Paul and Chen, 2020a]{paul2020random}
Paul, S. and Chen, Y. (2020a).
\newblock A random effects stochastic block model for joint community detection
  in multiple networks with applications to neuroimaging.
\newblock {\em Annals of Applied Statistics}, 14(2):993--1029.

\bibitem[Paul and Chen, 2020b]{paul2020spectral}
Paul, S. and Chen, Y. (2020b).
\newblock Spectral and matrix factorization methods for consistent community
  detection in multi-layer networks.
\newblock {\em The Annals of Statistics}, 48(1):230--–250.

\bibitem[Pensky, 2019]{pensky2019dynamic}
Pensky, M. (2019).
\newblock Dynamic network models and graphon estimation.
\newblock {\em The Annals of Statistics}, 47(4):2378--2403.

\bibitem[Pensky and Zhang, 2019]{pensky2019spectral}
Pensky, M. and Zhang, T. (2019).
\newblock Spectral clustering in the dynamic stochastic block model.
\newblock {\em Electronic Journal of Statistics}, 13(1):678--709.

\bibitem[Raskutti and Mahoney, 2016]{raskutti2016statistical}
Raskutti, G. and Mahoney, M.~W. (2016).
\newblock A statistical perspective on randomized sketching for ordinary
  least-squares.
\newblock {\em Journal of Machine Learning Research}, 17(1):7508--7538.

\bibitem[Rohe et~al., 2016]{rohe2016co}
Rohe, K., Qin, T., and Yu, B. (2016).
\newblock Co-clustering directed graphs to discover asymmetries and directional
  communities.
\newblock {\em Proceedings of the National Academy of Sciences},
  113(45):12679--12684.

\bibitem[Sakai and Imiya, 2009]{sakai2009fast}
Sakai, T. and Imiya, A. (2009).
\newblock Fast spectral clustering with random projection and sampling.
\newblock In {\em International Workshop on Machine Learning and Data Mining in
  Pattern Recognition}, pages 372--384. Springer.

\bibitem[Sinha, 2018]{sinha2018k}
Sinha, K. (2018).
\newblock K-means clustering using random matrix sparsification.
\newblock In {\em International Conference on Machine Learning}, pages
  4684--4692. PMLR.

\bibitem[Stark et~al., 2006]{stark2006biogrid}
Stark, C., Breitkreutz, B.-J., Reguly, T., Boucher, L., Breitkreutz, A., and
  Tyers, M. (2006).
\newblock Biogrid: a general repository for interaction datasets.
\newblock {\em Nucleic Acids Research}, 34(suppl\_1):D535--D539.

\bibitem[Su et~al., 2024]{su2024spectral}
Su, W., Guo, X., Chang, X., and Yang, Y. (2024).
\newblock Spectral co-clustering in multi-layer directed networks.
\newblock {\em Computational Statistics \& Data Analysis}, 198:107987.

\bibitem[Tremblay and Loukas, 2020]{tremblay2020approximating}
Tremblay, N. and Loukas, A. (2020).
\newblock Approximating spectral clustering via sampling: a review.
\newblock {\em Sampling Techniques for Supervised or Unsupervised Tasks}, pages
  129--183.

\bibitem[Tremblay et~al., 2016]{tremblay2016compressive}
Tremblay, N., Puy, G., Gribonval, R., and Vandergheynst, P. (2016).
\newblock Compressive spectral clustering.
\newblock In {\em International conference on machine learning}, pages
  1002--1011. PMLR.

\bibitem[Wang et~al., 2021]{wang2021fast}
Wang, J., Guo, J., and Liu, B. (2021).
\newblock A fast algorithm for integrative community detection of multi-layer
  networks.
\newblock {\em Stat}, 10(1):e348.

\bibitem[Wang et~al., 2019]{wang2019scalable}
Wang, S., Gittens, A., and Mahoney, M.~W. (2019).
\newblock Scalable kernel k-means clustering with {N}ystrom approximation:
  Relative-error bounds.
\newblock {\em Journal of Machine Learning Research}, 20(12):1--49.

\bibitem[Woodruff, 2014]{woodruff2014sketching}
Woodruff, D.~P. (2014).
\newblock Sketching as a tool for numerical linear algebra.
\newblock {\em Foundations and Trends{\textregistered} in Theoretical Computer
  Science}, 10(1--2):1--157.

\bibitem[Xie, 2024]{xie2024bias}
Xie, F. (2024).
\newblock Bias-corrected joint spectral embedding for multilayer networks with
  invariant subspace: Entrywise eigenvector perturbation and inference.
\newblock {\em arXiv preprint arXiv:2406.07849}.

\bibitem[Yan et~al., 2009]{yan2009fast}
Yan, D., Huang, L., and Jordan, M.~I. (2009).
\newblock Fast approximate spectral clustering.
\newblock In {\em Proceedings of the 15th ACM SIGKDD International Conference
  on Knowledge Discovery and Data Mining}, pages 907--916.

\bibitem[Zhang et~al., 2022]{zhang2022randomized}
Zhang, H., Guo, X., and Chang, X. (2022).
\newblock Randomized spectral clustering in large-scale stochastic block
  models.
\newblock {\em Journal of Computational and Graphical Statistics},
  31(3):887--906.

\end{thebibliography}
\end{document}